\documentclass[fleqn,usenatbib]{mnras}

\usepackage{newtxtext,newtxmath}

\usepackage[T1]{fontenc}

\DeclareRobustCommand{\VAN}[3]{#2}
\let\VANthebibliography\thebibliography
\def\thebibliography{\DeclareRobustCommand{\VAN}[3]{##3}\VANthebibliography}

\usepackage{graphicx}	
\usepackage{amsmath}	
\usepackage{pdflscape}

\usepackage{comment}

\newcommand{\Zsun}{${\rm Z}_{\odot}$}
\newcommand{\cii}{{\sc [CII]~}}

\newcommand{\oiii}{{\sc [OIII]~}}

\newcommand{\irxb}{${\rm IRX}$--$\beta$}
\newcommand{\irx}{${\rm IRX}$}
\newcommand{\muv}{${M_{\rm UV}}$}
\newcommand{\Lsun}{${\rm L}_{\odot}$}
\newcommand{\Msun}{${\rm M}_{\odot}$}
\newcommand{\Lir}{$L_{\rm IR}$}
\newcommand{\Luv}{$L_{\rm UV}$}

\newcommand{\mstar}{$M_{\star}$}

\title[REBELS-IFU SFRs]{REBELS-IFU: Steeply rising star formation histories and the importance of dust obscuration in massive $\mathbf{z \simeq7}$ galaxies revealed by multi-wavelength observations
}

\author[R. Fisher et al.]{R. Fisher,$^{1}$\thanks{E-mail: rebecca.fisher-7@postgrad.manchester.ac.uk (RF)}
R. A. A. Bowler,$^{1}$ 
R. K. Cochrane,$^{1}$ 
L. E. Rowland,$^{2}$
M. Stefanon,$^{3, 4}$
H. S. B. Algera,$^{5, 6, 7}$ \newauthor
M. Aravena,$^{8,9}$
R. Bouwens,$^{2}$ 
E. da Cunha,$^{10,11,12}$
P. Dayal,$^{13,14,15}$ 
A. Ferrara,$^{16}$
J. A. Hodge,$^{2}$
H. Inami,$^{6}$ \newauthor
L. Komarova,$^{3}$
R. Smit,$^{17}$
L. Sommovigo,$^{18}$
D. P. Stark,$^{19}$
P. P. van der Werf$^{2}$
\\
Affiliations are listed at the end of the paper
}

\date{Accepted XXX. Received YYY; in original form ZZZ}

\pubyear{2015}

\begin{document}
\label{firstpage}
\pagerange{\pageref{firstpage}--\pageref{lastpage}}
\maketitle

\begin{abstract}
Reliable star formation rate (SFR) measurements are essential for understanding early galaxy evolution, yet derived values rely on several assumptions. 
To address this problem, we investigate the SFRs of 12 massive ($9~<~\log$(\mstar/\Msun)~$<~10$) Lyman-break galaxies at $z=6.5-7.7$, drawn from the Atacama Large Millimeter/submillimeter Array (ALMA) Reionization Era Bright Emission Line Survey (REBELS) program. 
The multi-wavelength data, including \textit{JWST} NIRSpec IFU spectroscopy and ALMA observations, make this a unique sample for investigating SFR tracers at this epoch. 
We compare SFRs derived from the rest-UV, H$\alpha$, and far-infrared emission, and from spectral energy distribution (SED) fits.
We apply robust dust attenuation corrections, which are crucial since between $50-80$ per cent of the star formation is obscured, and find a stellar-to-nebular attenuation ratio of $f=0.50\pm0.08$, consistent with local star-forming galaxies.
The majority of the derived total SFRs (medians $25-120$ {\Msun}yr$^{-1}$) place the REBELS galaxies systematically above $z=7$ literature star-forming main-sequence relations, and our best-fit star formation histories (SFHs) rise more steeply than lower-mass galaxies at the same redshift.
We show that these rising SFHs mean commonly used luminosity-to-SFR conversion factors, derived assuming a constant SFH over given timescales, overestimate the SFRs averaged over these timescales for our galaxies.
We provide updated luminosity-to-SFR calibrations for $z\simeq7$ galaxies with rising SFHs, showing that commonly assumed rest-UV conversion factors overestimate the $100$~Myr average SFR by a factor of $\simeq3$.
Finally, we investigate burstiness indicators in the REBELS-IFU galaxies, finding that the rising SFHs imply that the H$\alpha$-to-UV luminosity ratio is an unreliable probe of bursty star formation.
\end{abstract}




\section{Introduction}
Measuring the cosmic star formation rate density (SFRD) provides key constraints on the efficiency at which galaxies form stars and evolve throughout the history of the Universe.
The SFRD is observed to rise rapidly to a peak around $z\simeq2$, and decline at later epochs \citep{Madau2014}.
Underpinning the reliability of SFRD estimates is the derivation of accurate star formation rates (SFRs) for individual galaxies, either from specific features in their observed spectra or from the star formation histories (SFHs) inferred from spectral energy distribution (SED) fits to the observations \citep[for reviews see][]{Kennicutt1998, Kennicutt2012, Madau2014}.
The availability of these features, or "tracers", depends on observational constraints, such as the wavelength coverage and sensitivity of the imaging or spectroscopic data obtained, in addition to the redshift and intrinsic properties of the galaxy.  
Since spectral features are not all produced by the same physical processes, obtaining consistent empirical calibrations for each tracer and making galaxy-to-galaxy SFR comparisons is non-trivial, with systematic offsets between SFRs derived via different methods for the same galaxies \citep[e.g.][]{Smit2016, Cochrane2021, Looser2023, Clarke2024, Pirie2024}.
It also remains unknown whether these calibrations, which are based on local sources, still hold at high redshift, where conditions in galaxies are expected to be more extreme with, for example, harder ionising radiation fields and higher electron densities in the interstellar medium \citep[ISM; e.g.][]{Cullen2016, Katz2023b, Roberts-Borsani2024, Topping2024c, Topping2025}.

The empirical calibrations used to obtain SFRs from observed luminosities are derived using stellar population synthesis models \citep[e.g.][]{Conroy2009, Bruzual2003}.  
The luminosity-to-SFR conversion rate factor for a given tracer, $\kappa_{\text{x}}$, can be calculated for a given stellar initial mass function (IMF), stellar mass range, SFH (usually constant), and stellar population synthesis model. 
In short, this involves combining stellar templates with a weighting determined by the IMF to create the synthetic spectra of single-age populations as a function of age.  With an assumed SFH, a full galaxy SED can then be produced by linearly combining these \citep[for more details see the reviews by][]{Kennicutt1998, Conroy2013}. 
Thus, the conversion factors derived from these models, at a minimum, depend on the SFH, galaxy age,  metallicity, and IMF, although the relative importance of each factor varies for different tracers and parameter regimes \citep[see][for details]{Madau2014}.
These models are needed since, while most of the stellar mass in young stellar populations resides in low mass stars, the massive stars emit the majority of the energy and, thus, most observational tracers probe the formation of massive stars. 
Different observables are also sensitive to slightly different stellar mass ranges \citep[for example, the stellar mass range probed by the rest-UV extends to lower masses than for H$\alpha$; ][]{Kennicutt2012, Madau2014}.
Thus, studying, for example, the SFR-stellar mass relation \citep[the star-forming main-sequence;][]{York2000, Brinchmann2004} as a function of redshift requires a careful, consistent combination of different tracers \citep[e.g.][]{Noeske2007, Whitaker2012, Whitaker2014, Speagle2014, Shivaei2015, Santini2017, Popesso2022, Koprowski2024}. 

The observations of galaxies at $z>6$ used to derive SFRs have typically been limited to the rest-UV, although the \textit{James Webb Space Telescope} (\textit{JWST}) has increased the accessibility of rest-optical wavelengths.
The rest-UV continuum of young star-forming galaxies in the range $1250-2500$~{\AA} is dominated by the light emitted by short-lived ($\lesssim100$ Myr), massive ($>3$~{\Msun}) stars \citep[e.g.][]{Huo2023}.  
The rest-UV luminosity is considered a good indicator of SFR under the assumption that significant fluctuations in the SFR occur over timescales longer than a few 10s of Myrs and is usually assumed to measure the average SFR over a timescale of around $\sim100$~Myr \citep{Hao2011}.  

SFRs can also be inferred from rest-optical emission lines, one of the most accessible and reliable of which is H$\alpha$ \citep[e.g.][]{Kennicutt1998, Moustakas2006}.  
H$\alpha$ emission primarily emanates from the recombination of gas in HII regions that have been photoionised by nearby, massive O stars that have short lifetimes \citep[$<20$~Myr; ][]{Kennicutt1998, Kennicutt2012, Madau2014}.
Thus, compared to the rest-UV emission, H$\alpha$ traces a narrower stellar mass range and shorter ($\sim10$~Myr) timescales \citep[e.g.][]{Murphy2011, Weisz2012, Faisst2019, Emami2019, Atek2022}.
Other lines such as Lyman $\alpha$, {\sc [OII]} $\lambda 3727$, and {\sc [OIII]} $\lambda 5007$ are also indicators of SFR, but the former is subject to complex radiative transfer effects and the latter two exhibit a complex dependency on the metallicity and excitation state of the ISM \citep[][]{Figueira2022}.

One of the key drawbacks of deriving SFRs from the rest-UV or H$\alpha$ is the impact of dust attenuation.
This cannot be ignored, even at high redshift, with studies suggesting that dust-obscured star formation could contribute of order 30 per cent of the total SFRD at $z=7$ \citep{Algera2022} and extrapolations suggesting the contribution could be around $20$ per cent at $z\simeq8$, and around $5$ per cent at $z\simeq10$ \citep{Liu2025, Zavala2021}.
Dust attenuation corrections are particularly important in massive galaxies ($\log(M_*/${\Msun}$) ={8.5-10}$), since these tend to have higher metallicities and thus dust content, with studies at $z=4-7$ finding nearly half of star formation in these massive galaxies to be dust-obscured \citep[e.g.][]{Dunlop2017, Bowler2018, Bowler2023, Fudamoto2021, Inami2022, Algera2023}. 

\begin{table*}
	\centering
	\caption{The 12 REBELS galaxies targeted with the NIRSpec IFU observations used in this study. Column (1): Galaxy identifier. Column (2): Spectroscopic redshift from the {\cii} detection from ALMA \citep{Bouwens2022}. 
    Column (3): Rest-frame UV-continuum slope, $\beta$, from \citet{Fisher2025}.
    Column (4): Balmer decrement, H$\alpha$/H$\beta$, derived from the emission line fluxes from \citet{Rowland2025} for the galaxies at $6.5\leq z < 7.0$ for which H$\alpha$ lies within the NIRSpec wavelength coverage. 
    Column (5):  Stellar mass, {\mstar}, from the BAGPIPES SED fits with a flexible dust attenuation curve \citep{Fisher2025}. Note that these differ from those presented in Stefanon et al. (in preparation), who assume a fixed Calzetti dust attenuation curve, but are consistent within the errors. 
    Column (6): Gas-phase metallicities, $Z_{\text{gas}}$, derived from optical emission lines \citep{Rowland2025}.
    Column (7): Colour excess for the ionised gas (nebular regions) derived using the Balmer decrement and Equation~\ref{eq:EBV_gas}, assuming the attenuation curves of \citet{Fisher2025}. 
    Column (8): Colour excess for stellar continuum calculated using Equation~\ref{eq:EBV_stellar}. 
    Column (9): Infrared-excess, {\irx}, using {\Lir} values from \citet{Bowler2023}, which assume a fixed dust temperature of $T_{\rm d} = 46$~K and emissivity index $\beta_{\rm d} = 2.0$, except for REBELS-25 and REBELS-38 (marked with $^*$) for which we use the {\Lir} values derived from multi-band ALMA data \citep{Algera2023, Algera2024}.
    }
	\label{tab:pt1}
	\begin{tabular}[]{cccccccccccr} 
        \hline
		ID & $z$ &$\beta$ & H$\alpha$/H$\beta$ & log($M_*$/\Msun) & 12+$\log$(O/H) & E(B-V)$_{\text{gas}}$ & E(B-V)$_{\text{stellar}}$ & IRX\\
        (1) & (2) & (3) & (4) & (5) & (6) & (7) & (8) & (9)\\
        \hline
        \hline
        REBELS-05 & 6.496 & $-1.42 \pm 0.06$ & $3.76 \pm 0.45$ & $9.60\substack{+0.10 \\ -0.10}$ & $8.51 \pm 0.16$ & $0.24 \pm 0.11$ & $0.11\substack{+0.03 \\ -0.03}$ & $0.58 \pm 0.09$ & \\
        REBELS-08 & 6.749 & $-1.92 \pm 0.05$ & $4.05 \pm 0.60$ & $9.31\substack{+0.10 \\ -0.10}$ & $8.22 \pm 0.20$ & $0.30 \pm 0.13$ & $0.08\substack{+0.02 \\ -0.02}$ & $0.59 \pm 0.09$ & \\
        REBELS-12 & 7.346 & $-1.67 \pm 0.03$ & -- & $9.80\substack{+0.09 \\ -0.09}$ & $8.23 \pm 0.13$ & -- & $0.06\substack{+0.03 \\ -0.03}$ & $0.39 \pm 0.12$ & \\
        REBELS-14 & 7.084 & $-1.74 \pm 0.03$ & -- & $9.54\substack{+0.14 \\ -0.12}$ & $7.90 \pm 0.12$ & -- & $0.06\substack{+0.03 \\ -0.02}$ & $0.23 \pm 0.10$ & \\
        REBELS-15 & 6.875 & $-2.01 \pm 0.03$ & $3.48 \pm 0.33$ & $9.40\substack{+0.03 \\ -0.03}$ & $7.78 \pm 0.30$ & $0.17 \pm 0.08$ & $0.09\substack{+0.02 \\ -0.02}$ & $<0.22$ & \\
        REBELS-18 & 7.675 & $-1.56 \pm 0.03$ & -- & $9.98\substack{+0.04 \\ -0.04}$ & $8.50 \pm 0.13$ & -- & $0.10\substack{+0.02 \\ -0.02}$ & $0.42 \pm 0.08$ & \\
        REBELS-25 & 7.307 & $-1.61 \pm 0.09$ & -- & $9.07\substack{+0.08 \\ -0.10}$ & $8.62 \pm 0.17$ & -- & $0.09\substack{+0.03 \\ -0.02}$ & $0.79\substack{+0.30 \\ -0.20}^*$ & \\
        REBELS-29 & 6.685 & $-1.89 \pm 0.05$ & $3.16 \pm 0.37$ & $9.94\substack{+0.08 \\ -0.06}$ & $8.73 \pm 0.15$ & $0.09 \pm 0.10$ & $0.09\substack{+0.03 \\ -0.02}$ & $0.32 \pm 0.10$ & \\
        REBELS-32 & 6.729 & $-1.34 \pm 0.07$ & $3.48 \pm 0.36$ & $9.75\substack{+0.13 \\ -0.11}$ & $8.48 \pm 0.13$ & $0.17 \pm 0.09$ & $0.12\substack{+0.04 \\ -0.04}$ & $0.67 \pm 0.13$ & \\
        REBELS-34 & 6.634 & $-2.23 \pm 0.03$ & $3.89 \pm 0.84$ & $9.59\substack{+0.08 \\ -0.09}$ & $8.33 \pm 0.29$ & $0.27 \pm 0.20$ & $0.03\substack{+0.02 \\ -0.01}$ & $<0.32$ & \\
        REBELS-38 & 6.577 & $-1.63 \pm 0.06$ & $3.98 \pm 0.54$ & $9.93\substack{+0.09 \\ -0.08}$ & $8.28 \pm 0.18$ & $0.28 \pm 0.12$ & $0.13\substack{+0.03 \\ -0.03}$ & $0.24\substack{+0.23 \\ -0.17}^*$ & \\
        REBELS-39 & 6.845 & $-2.07 \pm 0.04$ & $3.43 \pm 0.40$ & $9.57\substack{+0.10 \\ -0.11}$ & $8.02 \pm 0.29$ & $0.16 \pm 0.10$ & $0.05\substack{+0.02 \\ -0.01}$ & $0.30 \pm 0.09$ & \\
        \hline
	\end{tabular}
\end{table*}

One method to correct dust-attenuated fluxes involves assuming a dust attenuation curve, which describes the attenuation as a function of wavelength. 
Studies typically assume the attenuation or extinction curves derived from observations of the Milky Way or local galaxies \citep[e.g.][]{Cardelli1989, Calzetti2000, Gordon2003}, or infer the stellar and nebular attenuation curves from the sample of interest \citep[e.g.][]{Reddy2015, Reddy2020}.
An excess in attenuation towards nebular emission lines, known as differential reddening, has been observed in local star-forming galaxies \citep[e.g.][]{Fanelli1988, Calzetti1997, Calzetti2000, Kreckel2013, Koyama2018} and also at higher redshift \citep{ForsterSchreiber2009, Reddy2010, Kashino2013, Price2014, Reddy2015, Shivaei2020, Reddy2020, Barrufet2025}.
This can be explained by a physical picture of ionised gas being located around young, massive stars in birth clouds, which provides a larger dust-covering fraction and/or dust column density, whereas the stellar continuum can originate from a wider variety of stars across the galaxy \citep[e.g.][]{Calzetti1994, Charlot2000}.
This effect remains a key uncertainty when comparing SFR tracers \citep[e.g.][]{Faisst2019}.  
Some studies suggest that the continuum-to-nebular attenuation ratio (see Section~\ref{sec:results_dust_EBV}) may evolve from the $f=0.44$ factor found by \cite{Calzetti1997} to closer to $f=1$ by $z=2$ \citep[e.g.][]{Erb2006, Reddy2010, Kashino2013, Koyama2015, Valentino2015, Puglisi2016, Kashino2017, Faisst2019}.
The ratio of nebular to continuum attenuation remains poorly constrained at $z>2$, although measurements are beginning to be made with \textit{JWST} data \citep[e.g.][]{Tsujita2025}.

An alternative to correcting the rest-UV or optical emission using a dust attenuation curve is to combine it with an obscured SFR tracer that probes similar timescales, e.g. the rest-UV and far-infrared \citep[FIR, e.g.][]{Calzetti2012, Bouwens2012, Topping2022}.  
FIR SFRs are derived from the integrated luminosity between $8-1000~\mu$m \citep{Kennicutt1998}.  Dust absorbs the UV radiation from stars and re-emits it at longer wavelengths.
Therefore, the FIR luminosity is assumed to be proportional to the obscured SFR.  
This assumes dust heating from other sources, such as AGN and older stars, is negligible, which is likely a valid assumption at high redshift \citep{Madau2014}.
The FIR emission is typically modelled as a modified blackbody (MBB), which depends on the physical properties of the dust: the dust temperature ($T_{\rm d}$, which determines the wavelength at which the emission peaks), dust mass ($M_{\rm d}$), and dust emissivity index \citep[$\beta_{\rm d}$; e.g.][]{Casey2012, Sommovigo2020}.
The FIR luminosity-to-SFR conversion factor is obtained from models that compute the rest-UV and IR luminosities for various amounts of dust attenuation, with most assuming the foreground dust screen geometry of \cite{Calzetti2000}, although the exact model used for dust absorption is not thought to be significant \citep[][although see \citet{Sommovigo2025b}]{Madau2014}.
    
In this paper, we study a subsample of galaxies from the Reionisation Era Bright Emission Line Survey, known as the REBELS-IFU sample.  REBELS-IFU provides a unique dataset for comparing SFR tracers at $z\simeq7$ since multi-wavelength observations extending from the rest-frame UV/optical (\textit{JWST} NIRSpec spectroscopy) to the rest-frame FIR (ALMA) are available \citep[][Stefanon et al. in prep.]{Bouwens2022}.
This is the only sample of massive ({\mstar}~$>10^{9}$~{\Msun})  galaxies at $z>6$ that has this wealth of SFR tracers, providing a unique opportunity to investigate the consistency between different methods and shed light on the SFHs of these sources.  
As \textit{JWST} observations now probe increasingly high redshifts \citep[e.g.][]{Harikane2023}, where observations are typically limited to the rest-UV, it is important to investigate the reliability of rest-UV derived SFRs and the validity of assumptions, such as constant SFHs, in empirical calibrations.
From the NIRSpec spectra, we have measurements of key physical properties of these galaxies, including stellar mass (Stefanon et al. in prep.), dust attenuation curves \citep{Fisher2025}, and gas-phase metallicities \citep{Rowland2025}, allowing us to build up a comprehensive picture of these galaxies and reduce the number of assumptions required.

The structure of this paper is as follows:
In Section \ref{sec:data}, we introduce the REBELS-IFU sample and dataset.
In Section \ref{sec:methods}, we provide details of the methods we use to derive SFRs and apply dust attenuation corrections.
We present a detailed investigation of the dust attenuation in our galaxies in Section~\ref{sec:results}, which is necessary to accurately measure the obscured star formation.
Using these dust corrections, we derive SFRs from a range of different tracers in Section~\ref{sec:results_SFR_tracers}.
We show that our sources have significant dust-obscured star formation fractions and place our galaxies in the context of the star-forming main-sequence.
We discuss our results in Section~\ref{sec:discussion}, explaining discrepancies between SFRs by considering the effects of SFHs and tracer timescales.  We support our findings with simple model SEDs, which we use to investigate the effect of rising SFHs on luminosity-to-SFR conversion factors.
Finally, we use our results to discuss the reliability of the H$\alpha$-to-UV ratio as a diagnostic of SFH burstiness.  
The key findings are summarised in Section \ref{sec:summary}.  
We assume the standard $\Lambda$CDM cosmology with $H_0 = 70$ km s$^{-1}$ Mpc$^{-1}$, $\Omega_{\text{m}} = 0.3$, and $\Omega_{\Lambda} = 0.7$ throughout this work \citep{Planck2020}.

\section{Data and sample}
\label{sec:data}
The full details of the REBELS-IFU galaxies studied in this work are given in Stefanon et al. (in prep.) and \cite{Fisher2025}.  
In short, these galaxies are a subsample of the 40 rest-UV bright ({\muv}$<-21$) Lyman-break galaxies at $z=6.5-8$ with stellar masses $\log_{10}(M_*/${\Msun}$) = 8.8-10.4$ targeted by the REBELS ALMA large program \citep{Bouwens2022}.
Twelve galaxies with bright {\cii}$158 \mu$m emission in the REBELS program were selected for follow-up with \textit{JWST} NIRSpec observations (PID 1626; P.I. Stefanon and PID 2659; P.I. Weaver, \textit{JWST} Cycle 1).
These sources are listed in Table~\ref{tab:pt1}. 
The physical properties of these galaxies are representative of the full REBELS sample (see Stefanon et al. in prep.), but their SFRs inferred in previous work, for example from {\cii}of $50-400$~{\Msun}yr$^{-1}$, are significantly higher than typical galaxies at these redshifts \citep{DeLooze2014}.
The SFRs inferred from the rest-UV and FIR emission of the REBELS galaxies tend to be lower than those inferred from {\cii}\citep{Bowler2023, Ferrara2022, Sommovigo2022}, but still tend to place the REBELS sources systematically above literature star-forming main-sequence relations, which is unsurprising given their bright rest-UV selection.  However, the magnitude of the offset from the star-forming main-sequence depends strongly on the method used to derive stellar mass \citep{Topping2022, Algera2022}.

The NIRSpec integral field unit (IFU) observations probe the rest-UV and rest-optical wavelengths of the galaxies. These observations use the prism mode with a resolution of $R\simeq100$, and an exposure time of approximately 30 minutes per source.  
For each $0.08''$ spaxel in the $3.1''\times3.2''$ field of view, a spectrum covering the wavelength range $0.6-5.3$~$\mu$m is obtained.
In this work, we study the global properties of the galaxies derived from the integrated spectra \citep[see Stefanon et al. in prep. and][for details]{Rowland2025}.
The full details of the cube data reduction are presented in Stefanon et al. (in prep.), and we use the emission line flux measurements presented in \cite{Rowland2025}.

ALMA Band 6 or Band 7 observations of these sources were obtained as part of the Cycle 7 REBELS ALMA large program (PID: 2019.1.01634.L, P.I. Bouwens).
The dust continuum emission is simultaneously observed during the spectral scans searching for the {\cii}$158$~$\mu$m or {\oiii}$88$~$\mu$m lines.
All twelve of the REBELS-IFU galaxies were spectroscopically confirmed via the detection of their \cii $158 \mu$m line \citep{Bouwens2022}.
Dust continuum emission was also detected with a signal-to-noise ratio of $\geq 3.5 \sigma$ for ten of the twelve REBELS-IFU galaxies \citep[see Table~\ref{tab:pt1} and][]{Inami2022}.
Published multi-band ALMA observations are available for three sources and are used when these provide useful constraints on the FIR SED \citep[e.g.][]{Algera2023, Algera2024}.
Improved constraints on the remaining sources will soon be possible thanks to recent Band 8 observations (PID: 2024.1.00406.S, P.I. Algera). 

\begin{figure*}
\includegraphics[width=2\columnwidth]{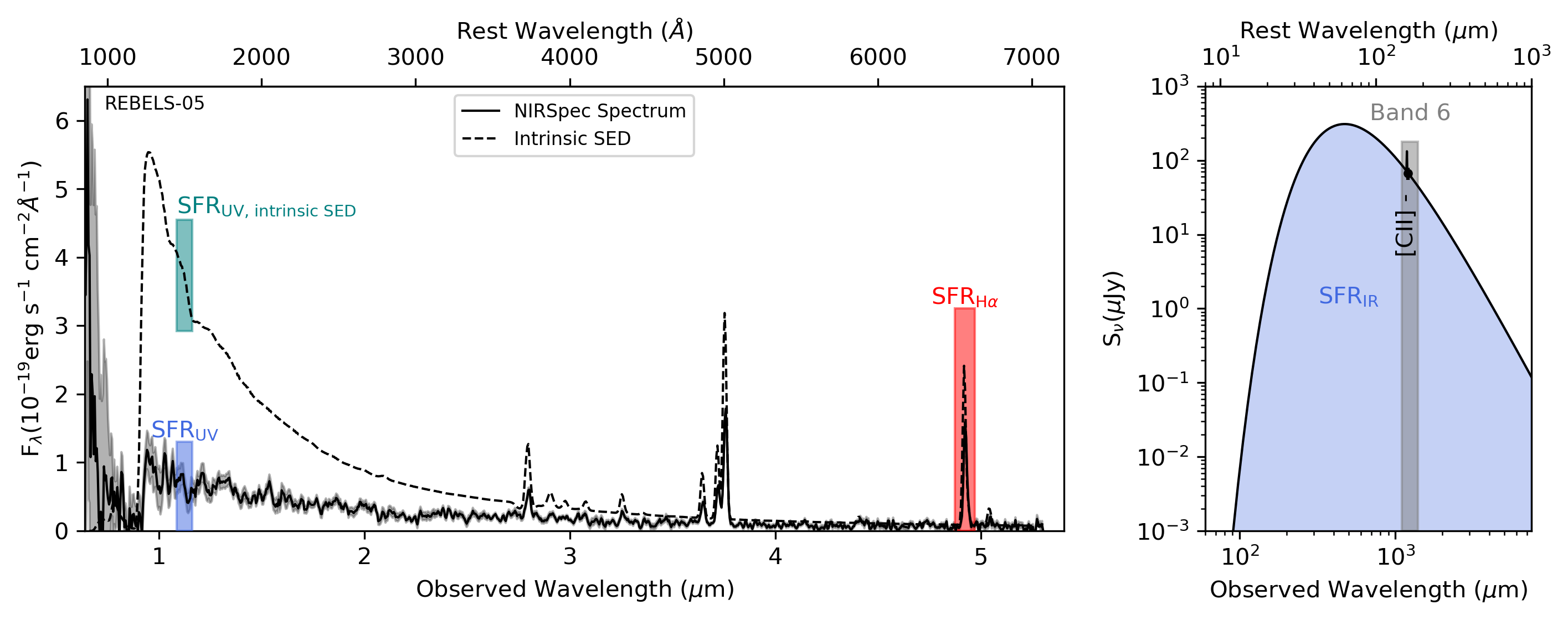}
\caption{We show with coloured rectangles the wavelength ranges used to infer star formation rates in this work.  The full NIRSpec PRISM spectrum for one of the REBELS-IFU galaxies (REBELS-05) is shown by the solid black line in the left panel.  The intrinsic, dust-free SED model is shown by the dashed line, demonstrating the wavelength-dependent effect of dust attenuation, with the rest-UV flux being impacted more than the H$\alpha$ flux.  The right panel shows a modified blackbody model of the rest-frame far-infrared emission. The ALMA Band 6 flux continuum measurement is shown by the black point.  The integrated flux under this curve traces the dust-obscured star formation rate.}
\label{fig:SFR_tracers_plot} 
\end{figure*}

\section{Methods}
\label{sec:methods} 

An illustration of the available data and the different methods that we use to infer SFRs is shown in Fig.~\ref{fig:SFR_tracers_plot}. 
We show an observed NIRSpec rest-UV/optical spectrum for one of our galaxies.  
The difference between this spectrum and the reconstructed intrinsic SED (dashed line) demonstrates the wavelength-dependent impact of dust attenuation on our SFR tracers, with the rest-UV being more attenuated than H$\alpha$.  
We will also make use of the non-parametric SFHs from the SED fits with a flexible dust attenuation curve to the full NIRSpec spectra presented in \cite{Fisher2025}.
In the right panel, we show a modified blackbody curve and the FIR ALMA dust continuum detection.
The shaded area under this curve is used to measure the dust-obscured SFR.   

\begin{table*}
	\centering
	\caption{The calibrations we assumed to convert luminosities to star formation rates for each tracer.}
	\label{tab:SFREqns}
	\begin{tabular}[]{lclcccr} 
		\hline
        SFR$_{\text{x}}$ & $=$ & $\kappa_{\text{x}} L_{\text{x}}$ & Source \\
        \hline
        \hline
        SFR$_{\text{UV}}$ & $=$ & $ 7.2 \times 10^{-29} L_{\nu}$ [erg s$^{-1}$ Hz$^{-1}$] & \cite{Madau2014} adjusted for \cite{Chabrier2003} IMF ($1.15 \times 10^{-28} \times 0.63$) \\
        SFR$_{\text{IR}}$  & $=$ & $1.1 \times 10^{-10} L_{\text{IR}}$ [\Lsun] & \cite{Madau2014} adjusted for \cite{Chabrier2003} IMF ($1.73 \times 10^{-10} \times 0.63$)\\
        
        SFR$_{\text{H}\alpha}$  & $=$ & $4.977 \times 10^{-42} L_{\text{H}\alpha}$ [erg s$^{-1}$] & \cite{Kennicutt1998} adjusted for \cite{Chabrier2003} IMF ($7.9 \times 10^{-42} \times 0.63$)\\
        \hline
	\end{tabular}
\end{table*}

\subsection{Deriving SFRs from the rest-UV, H$\alpha$, and FIR luminosities}
\label{sec:methods_lum_to_SFR}
The conversion factors, $\kappa_{\text{x}}$, we use in equations of the form SFR$_{\text{x}}~=~\kappa_{\text{x}}~L_{\text{x}}$ to obtain SFRs from the luminosities, $L_{\text{x}}$, of each tracer are given in Table~\ref{tab:SFREqns}.
For our fiducial values, we adopt those which are used extensively in the literature; however, we will discuss the effects of different metallicities and SFHs on these values in Section~\ref{sec:discussion_new_kappas}.
We adjust all the conversion factors to the \cite{Chabrier2003} IMF with stellar population mass limits of $0.1~-~100$~{\Msun} using the $0.63$ factor given in \cite{Madau2014}. 
For consistentcy with previous analysis of the REBELS sources \citep[e.g.][]{Bouwens2022, Algera2022, Bowler2023}, we adopt the standard conversion factors from \cite{Madau2014} for $\kappa_{\text{UV}}$ and $\kappa_{\text{IR}}$, which assume a constant SFR during the previous 100 Myr and a fixed stellar metallicity of $Z_*=0.1$~{\Zsun}.
The observed rest-UV luminosity, {\Luv}, was calculated from the spectrum flux, $F_{\lambda}$, in a top hat filter of width 100 {\AA} centred at $\lambda_{\text{rest}}=1500$ {\AA}. 
We use the rest-frame FIR luminosity values, {\Lir}, from \cite{Bowler2023}, which assume a dust temperature of $T_{\rm d} = 46$~K and a dust emissivity index of $\beta_{\rm d} = 2.0$, except for REBELS-25 and REBELS-38 for which we use the values derived by \cite{Algera2024} and \cite{Algera2023}, respectively, that use the multi-band ALMA observations available for these sources. 
This choice of $T_{\rm d}$ is justified based on the {\sc [CII]}-based method of \cite{Sommovigo2022}, and is consistent with the values obtained from the modified blackbody fits to multi-band ALMA observations \citep{Algera2023}.
We do not use the {\Lir} derived by \cite{Algera2023} for REBELS-12 since it is undetected in the Band 8 data and thus, the constraints on the FIR SED are poor. 
For SFR$_{\text{H}\alpha}$, we use the conversion factor from \cite{Kennicutt1998} which assumes an electron temperature of $T_e = 10^4$ K, solar stellar metallicity, Case B recombination, and a constant SFR for at least 10 Myr.
We take the H$\alpha$ emission line fluxes measured directly from the observed NIRSpec spectra from \cite{Rowland2025}. 

Here, we summarise some of the caveats and assumptions associated with these conversion factors, which will be discussed in more detail in Section~\ref{sec:discussion_SFH_model}.  
We note that each $\kappa$ value assumes a fixed stellar metallicity, whereas the gas-phase metallicities of our sample are known to vary between $Z_{\text{gas}} = 0.12-1.11$~{\Zsun} \citep[although these are not necessarily equal to the stellar metallicities; ][]{Rowland2025}.
$\kappa_{\text{H}\alpha}$ is more sensitive to metallicity than $\kappa_{\text{UV}}$, but our main conclusions remain unchanged even if we adopt other commonly used conversion factors or account for this variability.  

As shown in Fig.~\ref{fig:SFR_tracers_plot}, both the rest-frame UV and H$\alpha$ emission are attenuated by dust.  
To obtain total SFRs that include the contributions from dust-obscured star formation, an attenuated tracer can be combined with one that measures the obscured SFR over the same timescale, e.g. combining SFR$_{\text{UV}}$ and SFR$_{\text{IR}}$.
Alternatively, the total SFR can be obtained by applying a dust correction to convert the observed flux, $f_{\text{obs}}$, to the intrinsic flux,  $f_{\rm int}$, using
\begin{eqnarray}
     f_{\rm int}(\lambda) = f_{\text{obs}}(\lambda) 10^{0.4 A(\lambda)}
    \label{eq:fobs_fint},
\end{eqnarray}
where $A(\lambda)$ is the attenuation magnitude at that wavelength.
To correct the rest-UV (H$\alpha$) SFRs thus requires assuming a stellar (nebular) dust attenuation curve for each galaxy. 
We summarise our dust attenuation correction methods in the following subsection.

\subsection{Correcting SFRs with dust attenuation curves}
\label{sec:methods_dust_atten_curves}
The total dust attenuation curve, $k(\lambda)$, that describes the wavelength-dependent effect of dust absorption and scattering of stellar photons, is related to the attenuation magnitude, $A(\lambda)$, at wavelength $\lambda$ by
\begin{eqnarray}
    k(\lambda) = \frac{A(\lambda)}{E(B-V)}
    \label{eq:k_lambda_from_AlambdaAv},
\end{eqnarray} 
where $E(B-V)$ is known as the colour excess.
The colour excess is the difference between the B-band (4400~{\AA}) attenuation magnitude, $A_B$, and the V-band (5500~{\AA}) attenuation magnitude, $A_V$.
This is related to the total-to-selective extinction ratio, $R_V$, by 
\begin{eqnarray}
     E(B-V) = A_B - A_V = \frac{A_{V}}{R_V}
    \label{eq:fobs_fint}.
\end{eqnarray}

The colour excess of the ionised gas in a galaxy, $E(B-V)_{\text{gas}}$, can be calculated from the flux ratio of the hydrogen Balmer lines (the Balmer decrement) using 
\begin{eqnarray}
     E(B-V)_{\text{gas}} = \frac{2.5\log\left(\frac{\rm (H\alpha/H\beta)_{\text{obs}}}{\rm (H\alpha/H\beta)_{\text{int}}}\right)}{k(\lambda_{\rm H\beta})-k(\lambda_{\rm H\alpha})}
    \label{eq:EBV_gas},
\end{eqnarray}
where (H$\alpha$/H$\beta$)$_{\text{obs}}$ is the observed flux ratio and (H$\alpha$/H$\beta$)$_{\text{int}}$ is the intrinsic ratio.  
We assume Case B recombination with intrinsic ratios of H$\alpha$~:~H$\beta$~:~H$\gamma$~=~2.86~:~1.0~:~0.47 for a temperature of $10^4$ K and electron density $n_e=100$ cm$^{-3}$ \citep{Miller1974}.
It has been shown that this intrinsic ratio does not significantly vary for the range of physical conditions expected in regions of star formation \citep{Osterbrock2006, Smith2022}, although we note that anomalous ratios inconsistent with Case B recombination have recently been observed at high redshift \citep[e.g.][]{Yanagisawa2024}. 
For the four REBELS-IFU galaxies at $z>7$ (REBELS-12, REBELS-14, REBELS-18, and REBELS-25), H$\alpha$ is redshifted beyond the wavelength range of NIRSpec.
Since H$\gamma$ is blended with \oiii $\lambda 4363 + $[FeII] $\lambda 4360$ in the low-resolution prism spectra, we can only obtain a lower limit on the H$\beta$/H$\gamma$ ratio and thus cannot accurately dust correct the SFR$_{\text{H}\alpha}$ values (inferred from the H$\beta$ flux) for four of the REBELS-IFU galaxies \citep[for a detailed discussion see][]{Rowland2025}.

The colour excess of the stellar continuum, $E(B-V)_{\text{stellar}}$, is calculated using
\begin{eqnarray}
     E(B-V)_{\text{stellar}} = \frac{A_{V,\text{stellar}}}{R_V}
    \label{eq:EBV_stellar},
\end{eqnarray}
where $A_{V,\text{stellar}}$ and $R_V$ are obtained from the SED fits with a flexible dust attenuation curve presented in \citep{Fisher2025}.
For the \cite{Salim2018} power-law parameterisation of the dust attenuation curve slope, $R_V$ is calculated using
\begin{eqnarray}
     R_V = \frac{R_{V,\text{Calzetti}}}{(R_{V,\text{Calzetti}}+1)\left(4400/5500\right)^{\delta}-R_{V,\text{Calzetti}}}
    \label{eq:fobs_fint},
\end{eqnarray}
where $R_{V,\text{Calzetti}} = 4.05$ and $\delta$ is the deviation of the attenuation curve slope from the Calzetti-like curve.   

\subsection{Correcting SFRs with the IRX-$\mathbf{\beta}$ relation}
\label{sec:methods_IRX_beta}
The FIR-to-rest-UV luminosity ratio, known as the infrared-excess, {\irx}~$= \log_{10}($\Lir/\Luv), is a proxy for the obscured-to-unobscured SFR ratio.
Under the assumption of energy balance, the relation between this and the rest-frame UV slope, $\beta$, (the {\irxb} relation) provides indirect constraints on dust attenuation curves since {\Lir} is proportional to $M_{\rm d}$, whereas $\beta$ depends on the dust column density of the ISM along the line of sight to the observer.  
For the {\irxb} relation, we adopt the equation
\begin{eqnarray}
    {\rm IRX} = \log_{10} \left( 1.71 \times \left(10^{0.4 A(1600)} - 1 \right) \right)
    \label{eq:IRX_beta}
\end{eqnarray}
from \cite{McLure2018}.  Under the assumption that the intrinsic UV-slopes, $\beta_0$, of star-forming galaxies are similar, we can use the relation
\begin{eqnarray}
    A(1600) = \frac{{\rm d}A_{1600}}{{\rm d}\beta} (\beta - \beta_{0})
    \label{eq:IRX_beta_A1600},
\end{eqnarray}
where ${\rm d}A_{1600}/{\rm d}\beta$ is the slope of the dust attenuation curve.  For reference, for the \cite{Calzetti2000} dust attenuation curve ${\rm d}A_{1600}/{\rm d}\beta = 1.97$, while for the SMC extinction \citep{Gordon2003} ${\rm d}A_{1600}/{\rm d}\beta = 0.91$.
We measure the rest-frame UV-slope, $\beta$, directly from the observed spectra by fitting a power-law between the rest-frame wavelengths $\lambda_{\text{rest}}= 1268-2580$~{\AA} using a least-squares fitting method.
These values are consistent within the errors with fits using only the spectral windows defined by \cite{Calzetti1994}.  This suggests there are no strong emission or absorption features in the rest-UV region, with the largest difference in values being found in sources with a significant $2175$~{\AA} dust bump detection \citep{Fisher2025}.  
The relation 
\begin{eqnarray}
    M_{\rm UV} = -2.5\log(L_{\rm UV}) + 5.814
    \label{eq:Luv_in_Lsun}
\end{eqnarray}
was used to convert from the monochromatic rest-frame UV magnitude, {\muv}, to {\Luv} in units of {\Lsun} \citep{Capak2015}.  
Equation~\ref{eq:IRX_beta} can be fitted to data if rest-frame FIR observations are available, providing an indirect constraint on the average dust attenuation curve.
Alternatively, by assuming a dust attenuation curve and an intrinsic UV-slope, the obscured SFR of a galaxy can be inferred via the IRX, which is a proxy for SFR$_{\text{IR}}$/SFR$_{\text{UV}}$.

\section{Stellar and nebular dust attenuation corrections}
\label{sec:results}
In the following section, we present a detailed investigation of dust attenuation in our sources, which is necessary to accurately measure the contribution of dust-obscured star formation.

\begin{figure} 
\includegraphics[width=\columnwidth]{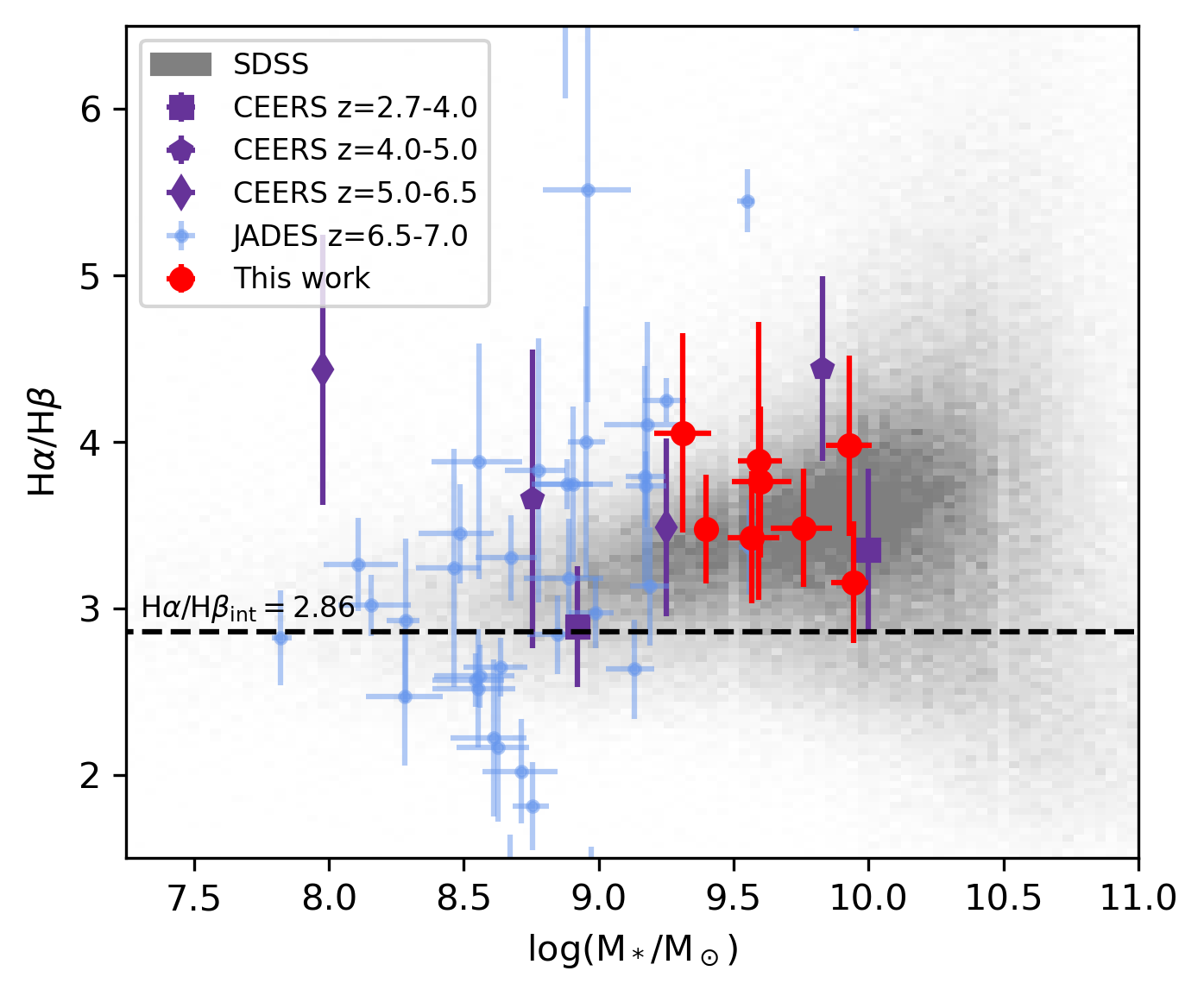}
\caption{The Balmer decrements, H$\alpha$/H$\beta$, are plotted against stellar mass, {\mstar}, in red for the 8 of the 12 massive REBELS-IFU galaxies at $6.5\leq~z~<~7.0$ for which H$\alpha$ lies within the NIRSpec wavelength coverage.  The H$\alpha$/H$\beta$ ratios all exceed the intrinsic Case B recombination value of 2.86, shown by the horizontal dashed line, indicative of dust attenuation affecting the nebular emission of these galaxies. 
The results from stacked CEERS galaxy spectra \citep{Shapley2023b} are shown in purple, and JADES galaxies at the same redshift as the REBELS-IFU sample are shown in blue.
The REBELS-IFU galaxies are consistent with local galaxies of similar stellar masses from the SDSS survey, shown by the grey histogram, suggesting no significant evolution in the Balmer decrement against stellar mass relation with redshift.
}
\label{fig:HaHb_vs_mass} 
\end{figure}

\subsection{The Balmer decrement}
\label{sec:results_dust_Balmer}
To measure the nebular dust attenuation and obtain dust-corrected H$\alpha$ SFRs, we use the Balmer decrement, H$\alpha$/H$\beta$. 
In Fig.~\ref{fig:HaHb_vs_mass}, we plot H$\alpha$/H$\beta$ against stellar mass for the eight REBELS-IFU galaxies at $6.5\leq z < 7.0$ for which H$\alpha$ lies within the NIRSpec wavelength coverage.   
The deviation of this ratio from the intrinsic value for Case B recombination of $2.86$ depends on the total mass of dust present and its spatial distribution in the galaxy.  
All of the REBELS-IFU H$\alpha$/H$\beta$ ratios exceed this intrinsic value, consistent with this nebular emission being dust attenuated. 

The REBELS-IFU galaxies are consistent with the positions of local, rest-optically selected SDSS galaxies with similar stellar masses, shown by the greyscale histogram \citep{Ahumada2020}.
This suggests that the dust properties of our sample, specifically the dust-mass surface density to stellar mass ratio \citep{Shapley2023b}, are not significantly different to local sources.
For comparison, we show the JADES galaxies in the same redshift range as our sample \citep{Eisenstein2023, Bunker2024, DEugenio2025} and results from the \textit{JWST} CEERS Survey using NIRSpec Micro-Shutter Assembly (MSA) stacked spectroscopy from \cite{Shapley2023b} spanning $z=2.7-6.5$. 
The REBELS-IFU galaxies lie at the higher redshift end of the sample from \cite{Shapley2023b} and the higher stellar mass end of the JADES galaxies at the same redshifts.  
The H$\alpha$/H$\beta$ ratios are consistent with galaxies of similar stellar masses across the full redshift range, supporting the findings of other studies that suggest that the Balmer decrement against stellar mass relation does not significantly evolve with redshift \citep[e.g.][]{Shapley2023b, Woodrum2025}.

\subsection{Differential dust attenuation is still important at $\mathbf{z\simeq7}$}
\label{sec:results_dust_EBV}

\begin{figure} 
\includegraphics[width=\columnwidth]{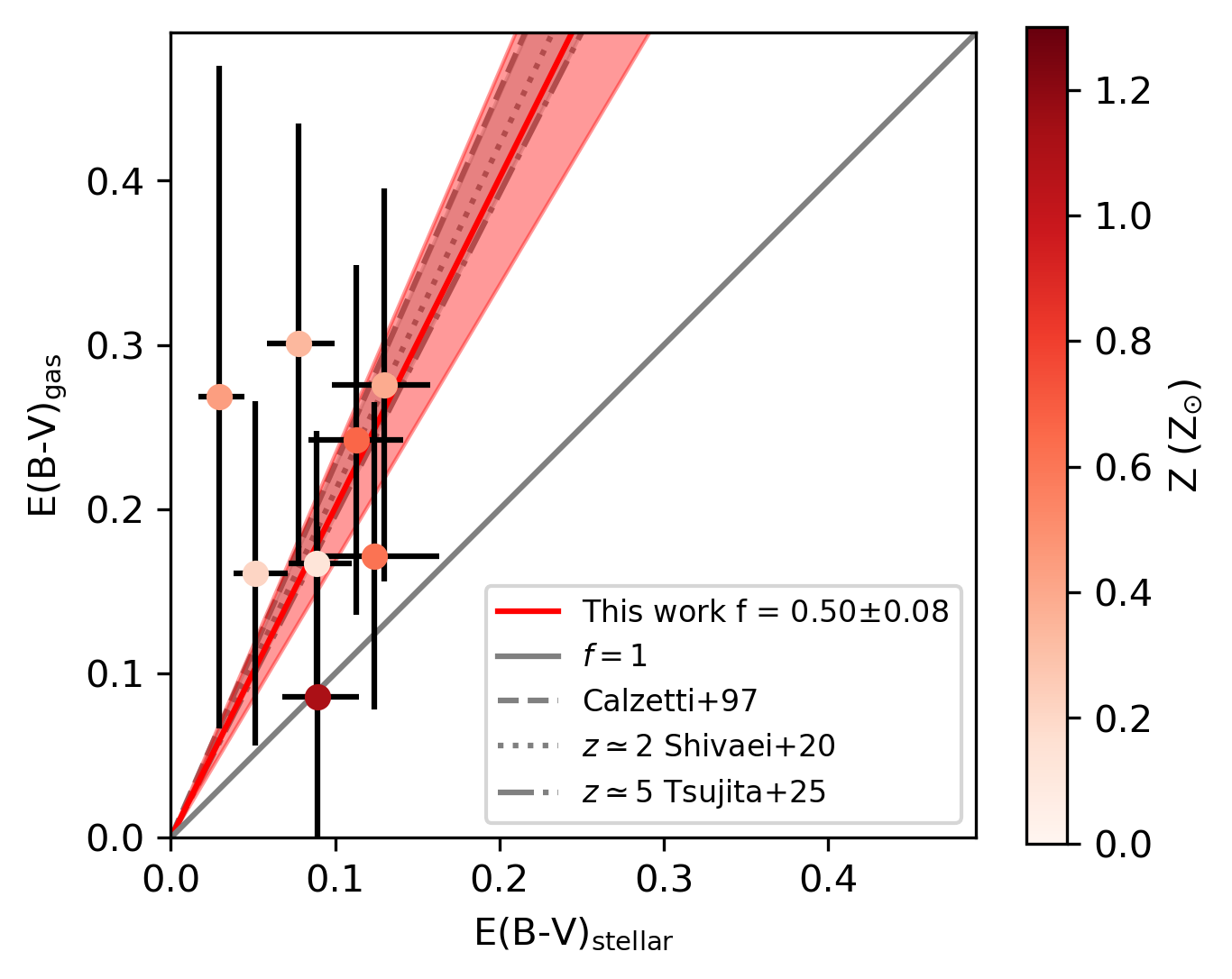}
\caption{
The colour excess measured for the ionised gas, $E(B-V)_{\text{gas}}$, derived from the Balmer decrement, compared to the colour excess measured for the stellar continuum, $E(B-V)_{\text{stellar}} = A_{V, \text{stellar}}/{R_V}$, for the 8 of the 12 massive galaxies in the REBELS-IFU sample at $6.5\leq z < 7.0$ for which H$\alpha$ lies within the NIRSpec wavelength coverage.
The average stellar-to-nebular attenuation ratio of $f = (0.50 \pm 0.08)$ is shown by the red line, indicating on average the nebular regions in these galaxies are approximately $2$ times more dust-obscured than the stellar continuum.
This is consistent with the $f = 0.44$ relation (dashed line) derived for local star-forming galaxies by \citet{Calzetti1997b}, and the relations for the ALPINE galaxies at $z\simeq5$ \citep{Tsujita2025} and the MOSDEF galaxies at $z\simeq2$ \citep{Shivaei2020}, shown by the dash-dot and dotted lines, respectively.
The points are coloured by their gas-phase metallicities \citep{Rowland2025}.
}
\label{fig:EBV_gas_vs_stellar} 
\end{figure}

In Fig.~\ref{fig:EBV_gas_vs_stellar} we compare the colour excess determined from the ionised gas (nebular regions) derived from H$\alpha$/H$\beta$ (Equation~\ref{eq:EBV_gas}) to those for the stellar continuum (Equation~\ref{eq:EBV_stellar}).
In most cases, the colour excess of the ionised gas exceeds that measured for the stellar continuum.
The linear fit to the eight REBELS-IFU galaxies for which H$\alpha$ lies within the NIRSpec wavelength coverage gives a stellar-to-nebular attenuation ratio $f = E(B-V)_{\text{stellar}}/E(B-V)_{\text{gas}}=~0.50\pm0.08$.
This is consistent, within the errors, with the $f = 0.44$ factor derived by \cite{Calzetti1997b} for local star-forming galaxies.
This value also shows good agreement with the value of $f=0.51\substack{+0.04 \\ -0.03}$ derived for the main-sequence ALPINE galaxies at $z\simeq5$ \citep{Tsujita2025, Faisst2025b} and the relation seen in star-forming galaxies at $z=1.4-2.6$ by \cite{Shivaei2020}.
Thus, we see no clear evolution in the ratio with redshift \citep[in contrast to recent results from the JADES survey;][]{Woodrum2025}.
However, we note that there is a large degree of scatter around this relation for individual galaxies, as has also been seen in other studies \citep[e.g.][]{Reddy2015, Reddy2020, Theios2019, Shivaei2020, Atek2022}, and that the errors on individual $f$ values remain large even with our strong detections of the Balmer lines.
While our average relation is consistent with the commonly assumed $f=0.44$ factor, we caution that for individual galaxies, inferring the nebular attenuation from the continuum attenuation using this linear relation is unreliable since $f$ ranges between $0.1$ and $1$ in our sample (with a standard deviation of $0.27$).
This justifies our decision to exclude from our analysis the SFR$_{\text{H}\alpha}$ values for galaxies in the REBELS-IFU sample for which we cannot measure the Balmer decrement.

One of the key strengths of our comparison is that we use the stellar attenuation curve derived for each galaxy in \cite{Fisher2025}.
However, we note that it is often assumed that the nebular and stellar attenuation curves are different, with the \cite{Cardelli1989} curve often assumed for the nebular regions.
Using a sample of the MOSDEF galaxies, \cite{Reddy2020} measured the nebular attenuation curve at $z\simeq2$ and did indeed find a relation close to the \cite{Cardelli1989} Galactic extinction curve at rest-optical wavelengths. 
In local SDSS galaxies, \cite{Rezaee2021} finds the nebular attenuation curve has a near-universal shape, similar in both shape and normalisation to the Galactic \citep{Cardelli1989}, SMC \citep{Gordon2003}, and MOSDEF results at $z\simeq2$ \citep{Reddy2020} within $2\sigma$, $2\sigma$, and $1\sigma$, respectively.  
However, the nebular attenuation curve measured for a galaxy at $z=4.41$ deviates significantly from local templates \citep{Sanders2024b}.  
Unfortunately, we are unable to place strong constraints on the nebular attenuation curve for our galaxies by comparing multiple Balmer line ratios, since in our PRISM spectra H$\gamma$ is blended with other emission lines and the signal-to-noise (S/N) ratios of higher-order Balmer emission lines are too low. 
Thus, the nebular attenuation curve remains a key uncertainty at high redshift.
We find very tentative suggestions of deviations from standard curves at shorter wavelengths when using a stack of the spectra, but investigating this further in our sample requires higher resolution spectroscopy to deblend the Balmer lines, with sufficient depth to enable high S/N measurements of the higher-order lines. 
In Appendix~\ref{sec:EBV_different_neb_curve}, we explore the impact of different assumptions about nebular and stellar attenuation curves on the inferred $f$ values. The inferred $f$ varies slightly with different assumptions (e.g. $f = (0.50 \pm 0.08)$ for our fiducial setup, compared to $f = (0.42 \pm 0.07)$ when assuming a Cardelli extinction curve for the nebular component, and $f = (0.56 \pm 0.10)$ when assuming a Calzetti curve for both nebular and stellar components).
However, these values are formally consistent at the 1-sigma level.
Thus, our assumption throughout this paper that the dust attenuation curves derived in \cite{Fisher2025} apply to both the ionised gas and stellar continuum does not significantly affect our conclusions.

\begin{figure} 
\includegraphics[width=\columnwidth]{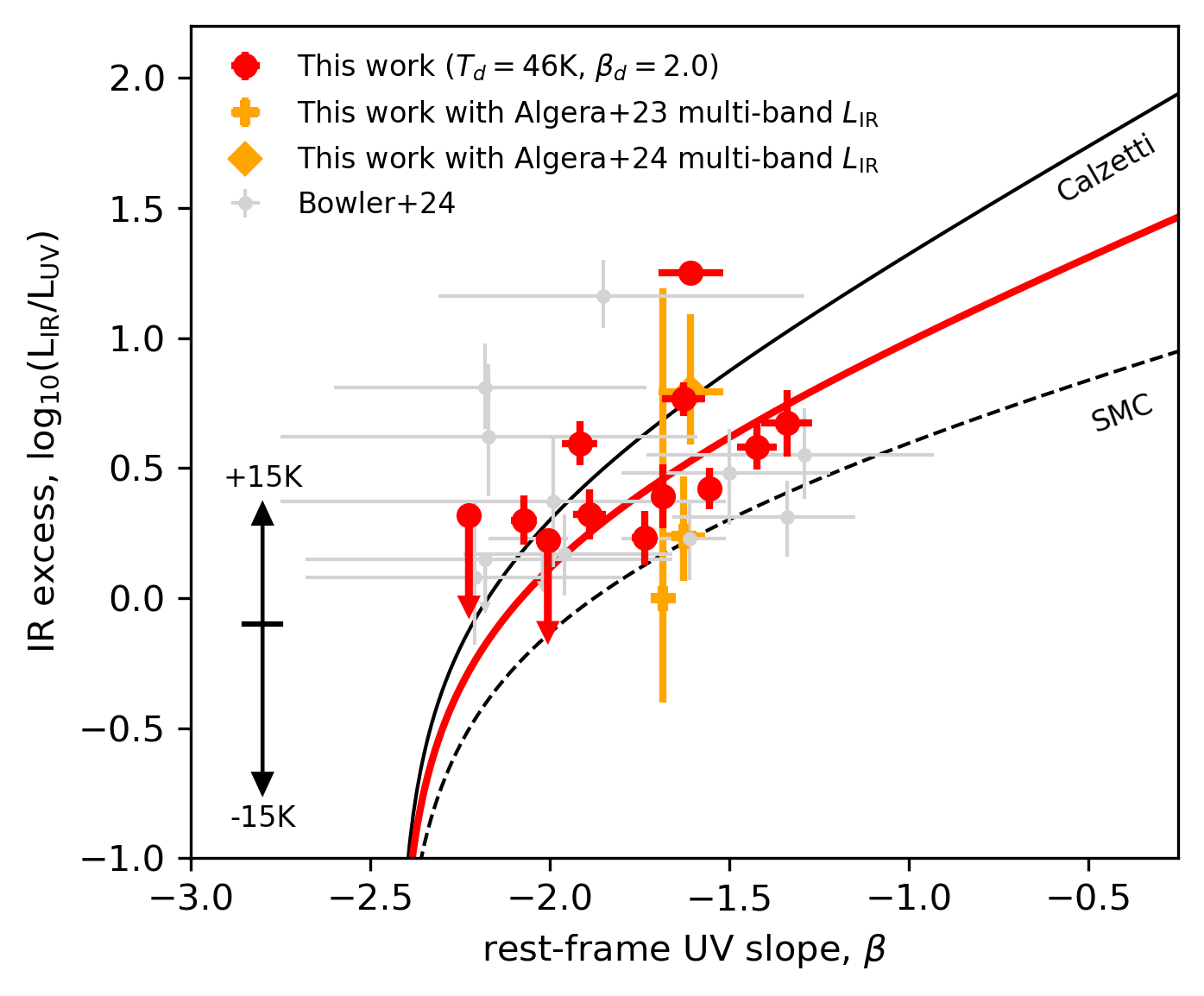}
\caption{The {\irxb} points for the 12 REBELS-IFU galaxies at $z\simeq7$.
The rest-frame UV continuum slope, $\beta$, and luminosity, {\Luv}, values are measured directly from the NIRSpec spectra.  In red we show the points using {\Lir} derived by \citet{Bowler2023}, which assume $T_{\rm d} = 46$ K and $\beta_{\rm d} = 2.0$, but with rest-UV quantities re-derived from the NIRSpec spectra.
The orange points use the {\Lir} values of \citet{Algera2023} and \citet{Algera2024} derived from multi-band ALMA observations.
The red line shows the best-fit {\irxb} relation with the intrinsic rest-UV slope fixed to the median value from \citet{Fisher2025}: $\beta_{0} = -2.43$.
This lies between the expected relations for the Calzetti-like attenuation and SMC extinction curves, shown by the solid and dashed black lines, respectively.
}
\label{fig:IRX} 
\end{figure}

We find tentative suggestions of a correlation between the ratio of the gas-to-stellar attenuation and the gas-phase metallicity, with the most metal-rich galaxy (REBELS-29) lying on the 1-to-1 relation. 
However, we caution that the correlation is not statistically significant and there are large error bars on $E(B-V)_{\text{gas}}$.
A trend between this ratio and metallicity has been seen at $z\simeq2$ in the MOSDEF galaxies and is consistent with the findings of other studies \citep[see][and references therein]{Shivaei2020}.
This can be explained physically if in high metallicity galaxies there is a high surface density of dusty clouds that affect both the ionised gas and stellar continuum, whereas in lower metallicity galaxies there are two distinct dust components: the diffuse ISM dust and dusty birth clouds.  
\cite{Pannella2015} suggest that galaxies with high star formation surface densities and larger dust-to-gas (DtG) ratios have dense ISMs, which reddens the stellar continuum to a similar degree as the nebular emission.
If our galaxies followed the \cite{RemyRuyer2014} local relation of increasing DtG ratios with metallicity, this would support this interpretation \citep[see][]{Algera2025}.

\subsection{The IRX-$\mathbf{\beta}$ relation}

ALMA observations of the rest-frame FIR emission provide measurements of the infrared luminosity, a proxy for the dust-obscured SFR. 
In Fig.~\ref{fig:IRX} we show the {\irxb} relation using the $\beta$ and {\Luv} values obtained from the NIRSpec spectra in combination with {\Lir} values derived by \cite{Bowler2023}, which assume $T_{\rm d} = 46$~K and $\beta_{\rm d} = 2.0$.
We also show the values using {\Lir} derived from multi-band ALMA data where these are available \citep{Algera2023, Algera2024}.
These have larger errors since the modified blackbody fits do not assume a fixed dust temperature and thus have more free parameters. 
The errors on the rest-UV slope and {\Luv} are significantly reduced compared to the photometrically derived values from \cite{Bowler2023}.
The scatter in $\beta$ values is also significantly reduced.

The {\irxb} points shown in Fig.~\ref{fig:IRX} mostly lie between the \cite{Calzetti2000} starburst relation and the SMC extinction curve.
We fit the data points of the dust continuum-detected galaxies, replacing the {\Lir} values for REBELS-25 and REBELS-38 with those presented in \cite{Algera2024} and \cite{Algera2023}, respectively, since these are well-constrained by multi-band ALMA observations.
We fix $\beta_{0} = -2.43$; the median intrinsic UV-slope from the SED models of \cite{Fisher2025} and obtain ${dA_{1600}}/{d\beta} = 1.44 \pm 0.14 $.
The fit is consistent with the average attenuation curve slope found in \cite{Fisher2025}, which is slightly steeper than the \cite{Calzetti2000} attenuation curve.
This result is also consistent with the value obtained by \cite{Bowler2023} using the full REBELS sample of ${dA_{1600}}/{d\beta} = 1.38 \pm 0.09$ when assuming $\beta_0=-2.5$. 

We note the usual caveats of assuming a fixed dust temperature, $T_{\rm d}$, when calculating {\Lir}.  With a fixed dust emissivity index, a change of temperature of 10 K can shift {\irx} values by 0.4 dex \citep{Mitsuhashi2023a, Bowler2023}.  
For most of our sample, only a single ALMA band detection is available to constrain the modified blackbody SED fit to the rest-frame FIR continuum (although recently obtained observations will increase this to two in the future; Algera et al. in prep.).  
The source with the best constrained FIR SED in our sample, REBELS-25, has been observed in six ALMA bands (5 detections, 1 limit), and the FIR SED fit has yielded a $T_{\rm d} = 32^{+9}_{-7}$ and $\beta_{\rm d} = 2.5\pm0.4$ \citep{Algera2024}.
As shown in Fig.~\ref{fig:IRX}, this causes a downwards vertical shift in the position of this galaxy compared to assuming $T_{\rm d} = 46$~K.
Thus, the average attenuation curve slope obtained from the independent methods of fitting directly to the NIRSpec spectra \citep{Fisher2025} and using the {\irxb} relation are consistent, suggesting that a stellar attenuation curve between the Calzetti-like and SMC relations is most appropriate for the REBELS galaxies.
We therefore assume the attenuation curves of \cite{Fisher2025} for dust correcting our SFRs in the following sections.

\section{Star Formation Rates and the star-forming main-sequence}
\label{sec:results_SFR_tracers}
With these methods to measure dust-obscured star formation, we now compare the SFRs of the REBELS-IFU galaxies inferred using different tracers.
We define the SFRs listed in Table~\ref{tab:pt2} as follows:
\begin{itemize}
    \item SFR$_{\text{UV}}$ - derived from the rest-UV luminosity of the observed NIRSpec spectra at rest-frame $1500$~{\AA} using a top-hat filter of width $100$~{\AA}, uncorrected for dust attenuation.
    \item SFR$_{\text{UV, intrinsic SED}}$ - derived from the reconstructed intrinsic (dust-free) SED model \citep[i.e. corrected for dust attenuation using the curves derived from fits to the NIRSpec spectra in][]{Fisher2025} at rest-frame $1500$~{\AA} using a top-hat filter of width $100$~{\AA}.
    \item SFR$_{\text{H}\alpha}$ - derived from the H$\alpha$ emission line flux, corrected for dust attenuation using the Balmer decrement, assuming the attenuation curve derived for each galaxy in \cite{Fisher2025}.
    \item SFR$_{\text{IR}}$ - dust-obscured SFR derived from the rest-frame FIR ALMA dust continuum detection.  We assume a dust temperature of $46$~K \citep{Bowler2023}, except for REBELS-25 and REBELS-38, where we use the FIR luminosities presented in \cite{Algera2024, Algera2023} derived from multi-band ALMA observations.
    \item SFR$_{\text{IR, IRX}}$ - dust-obscured SFR inferred from the observed rest-UV luminosity and the rest-UV continuum slope from the NIRSpec spectra assuming the {\irxb} relation fitted in Fig.~\ref{fig:IRX}.
    \item SFR$_{10~\text{Myr}}$ or SFR$_{100~\text{Myr}}$ - SFR averaged over $10$~Myr or $100$~Myr, inferred from the non-parametric SFH of the SED model fitted to the NIRSpec spectra \citep[the same SED model from which SFR$_{\text{UV, intrinsic SED}}$ is derived; see][for details of the SED fitting]{Fisher2025}.  These SFRs use all the information in the full NIRSpec spectra rather than a limited wavelength range as in the case of the rest-UV/H$\alpha$ SFRs.
    These SED fits do not use the low S/N ALMA FIR data, meaning very obscured regions could potentially be unaccounted for \citep[e.g.][]{Alvarez-Marquez2023}. 
    However, as discussed in Section 5.3 of \cite{Fisher2025}, values derived from the SED fits (e.g. $A_V$) show good consistency with FIR-derived properties, suggesting this is not the case.
    \item SFR$_{\text{[CII]}~158~\mu\text{m}}$ - SFR inferred from the FIR {\cii}$158~\mu$m emission line flux, using the starburst calibration from \cite{DeLooze2014}.  
\end{itemize}

\begin{figure} 
\includegraphics[width=\columnwidth]{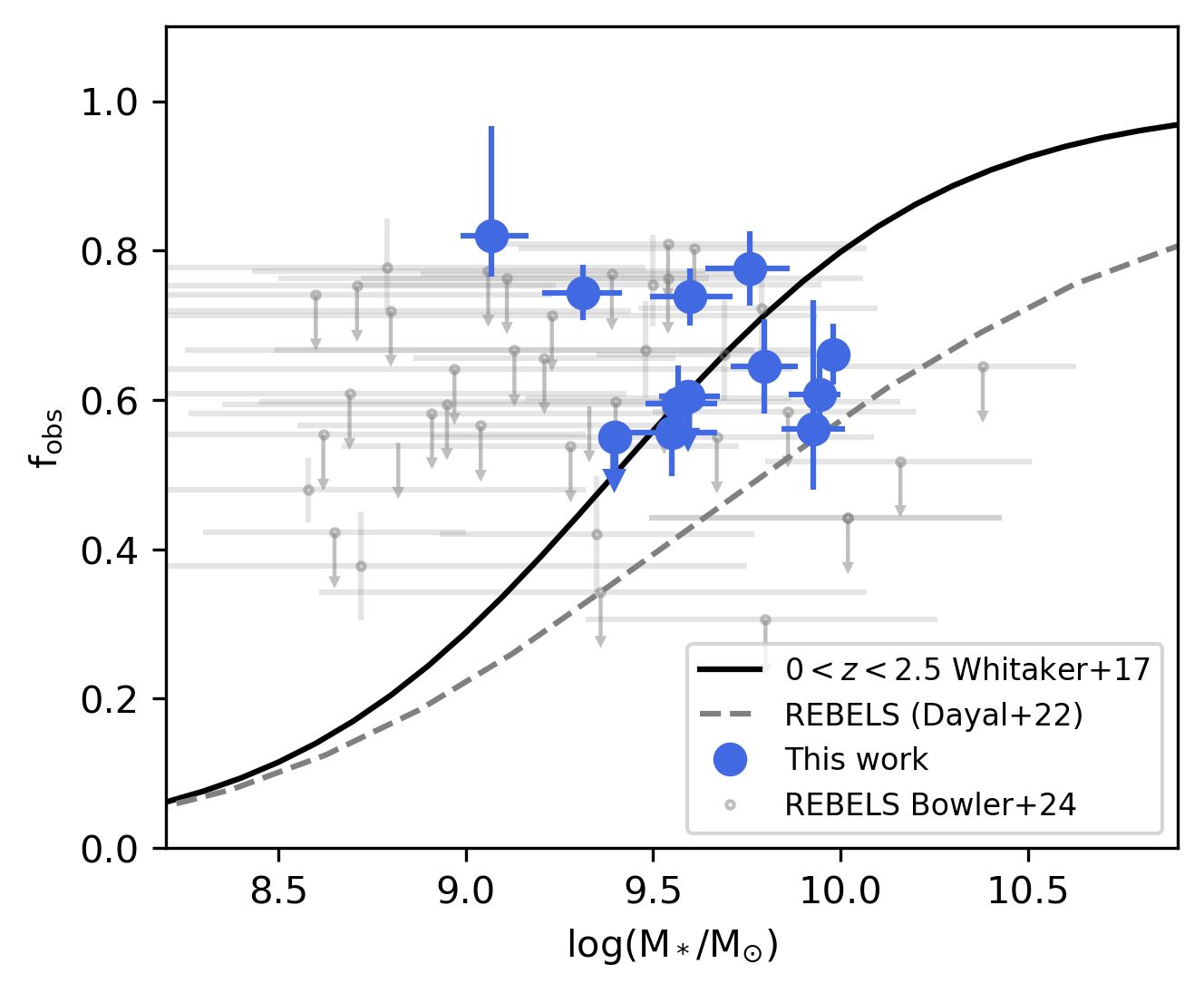}
\caption{The obscured SFR fractions, $f_{\text{obs}} =$~SFR$_{\text{IR}}$/SFR$_{\text{UV+IR}}$, versus stellar mass for the 12 REBELS-IFU galaxies at $z\simeq7$ (blue).
The REBELS-IFU galaxies have significant dust-obscured SFR fractions, demonstrating the necessity of robust dust corrections.
The small grey points show the obscured fractions for the other REBELS galaxies from \citet{Bowler2023}.
The $f_{\text{obs}}$--{\mstar} relation derived from a mass-complete sample of galaxies at $0<z<2.5$ by \citet{Whitaker2017} is shown by the black line.  
}
\label{fig:fobs} 
\end{figure}

\subsection{Significant dust-obscured star formation fractions}

To demonstrate the importance of dust-obscured star formation in our sources, in Fig.~\ref{fig:fobs} we show the dust-obscured SFR fractions calculated from $f_{\text{obs}} = $~SFR$_{\text{IR}}$/SFR$_{\text{UV+IR}}$ (where SFR$_{\text{UV+IR}}=$SFR$_{\text{UV}}+$SFR$_{\text{IR}}$).  
We find high obscured fractions, in the range $f_{\text{obs}} = 0.56-0.78$, consistent with those previously derived for the REBELS sources from ground-based photometry \citep{Inami2022, Bowler2023, Dayal2022, Ferrara2022, Sommovigo2022}, with a significant reduction in the stellar mass uncertainties compared to the previous studies.  
This is consistent with previous results that nearly half of the star formation in massive ($\log$(\mstar/\Msun)$ > 9$) galaxies at high-redshift ($z\gtrsim4$) is known to be dust obscured \citep[e.g.][]{Bowler2018, Bowler2023, Fudamoto2021, Inami2022, Schouws2022,  Algera2023, Barrufet2023}.
For comparison, we show the $f_{\text{obs}}$--{\mstar} relation derived from a mass-complete sample of galaxies at $0<z<2.5$ by \cite{Whitaker2017}.  
The limited stellar mass range and selection effects of our sample prevent us from ascertaining whether this relation holds at these redshifts; however, the majority of the REBELS-IFU galaxies are approximately consistent with this relation.
The four galaxies with the highest $f_{\text{obs}}$ values are offset above the \cite{Whitaker2017} relation.  
This could reflect intrinsic scatter, or it might suggest that the SED-derived stellar masses are slightly underestimated due to the rest-UV/optical spectra being dominated by the less obscured regions  \citep[there is evidence for spatial segregation between the dust and rest-UV emitting regions in these galaxies, see][]{Inami2022}. 
Nevertheless, the high obscured star formation fractions highlight the necessity of robust dust corrections when comparing the total SFRs for our massive sources, which are more significant than in lower stellar mass sources. 
We also plot the theoretical predictions based on the REBELS sample from \cite{Dayal2022} for comparison, which uses the SFRs calculated in \cite{Ferrara2022}.
The obscured star formation rate fractions derived from H$\alpha$, $f_{\text{obs, H}\alpha}=1 - (\text{SFR}_{\text{H}\alpha, \text{uncorrected}}/\text{SFR}_{\text{H}\alpha})$, are lower due to the wavelength-dependence of dust attenuation, but are still significant, ranging between $0.2$ and $0.6$, demonstrating the need to understand both nebular and continuum dust attenuation (see Section~\ref{sec:results}).

\subsection{Consistency between SFRs derived from the rest-UV}
\label{sec:results_rest_UV_tracers}

\begin{figure*} 
    \includegraphics[width=\columnwidth]{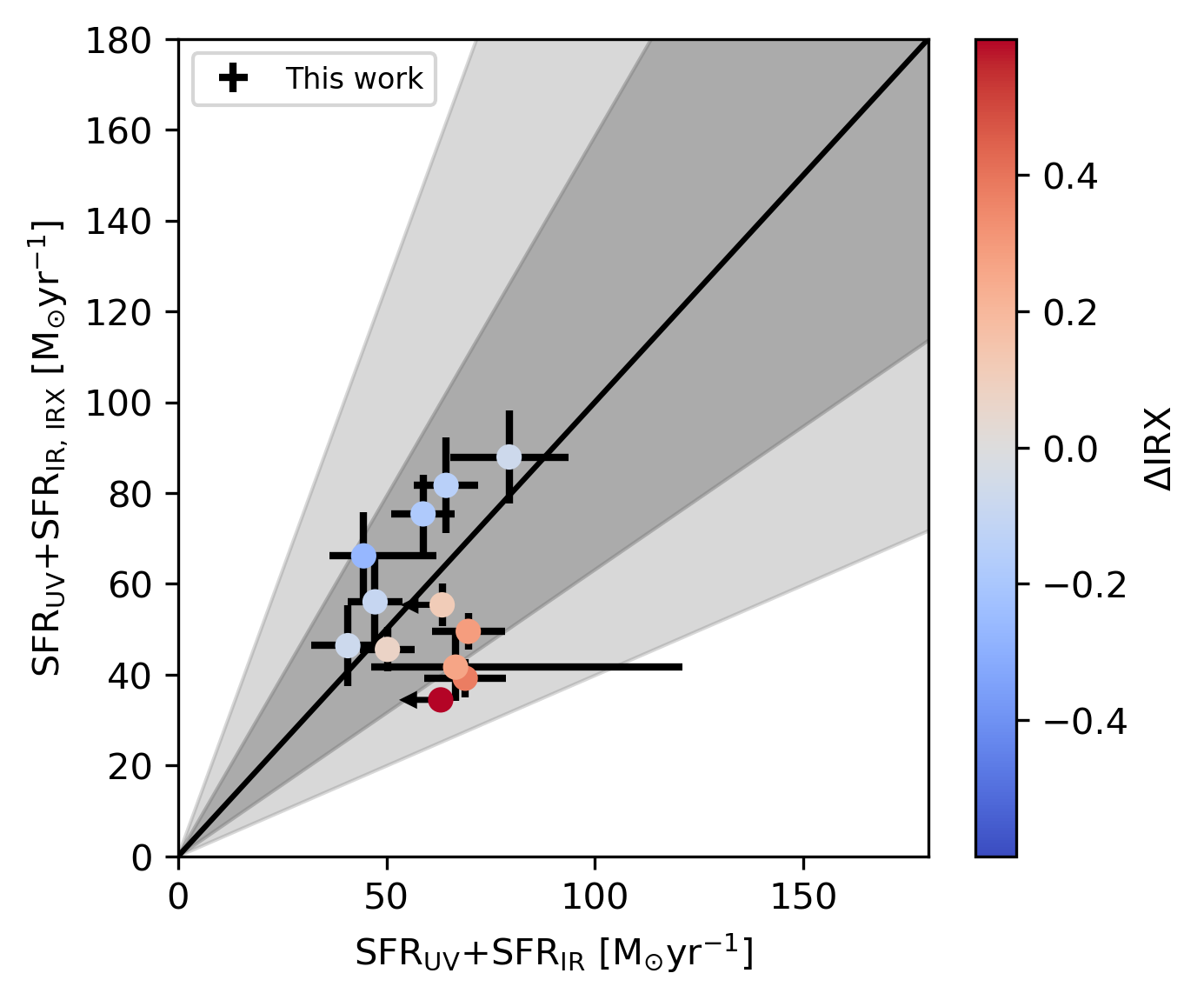}
    \includegraphics[width=\columnwidth]{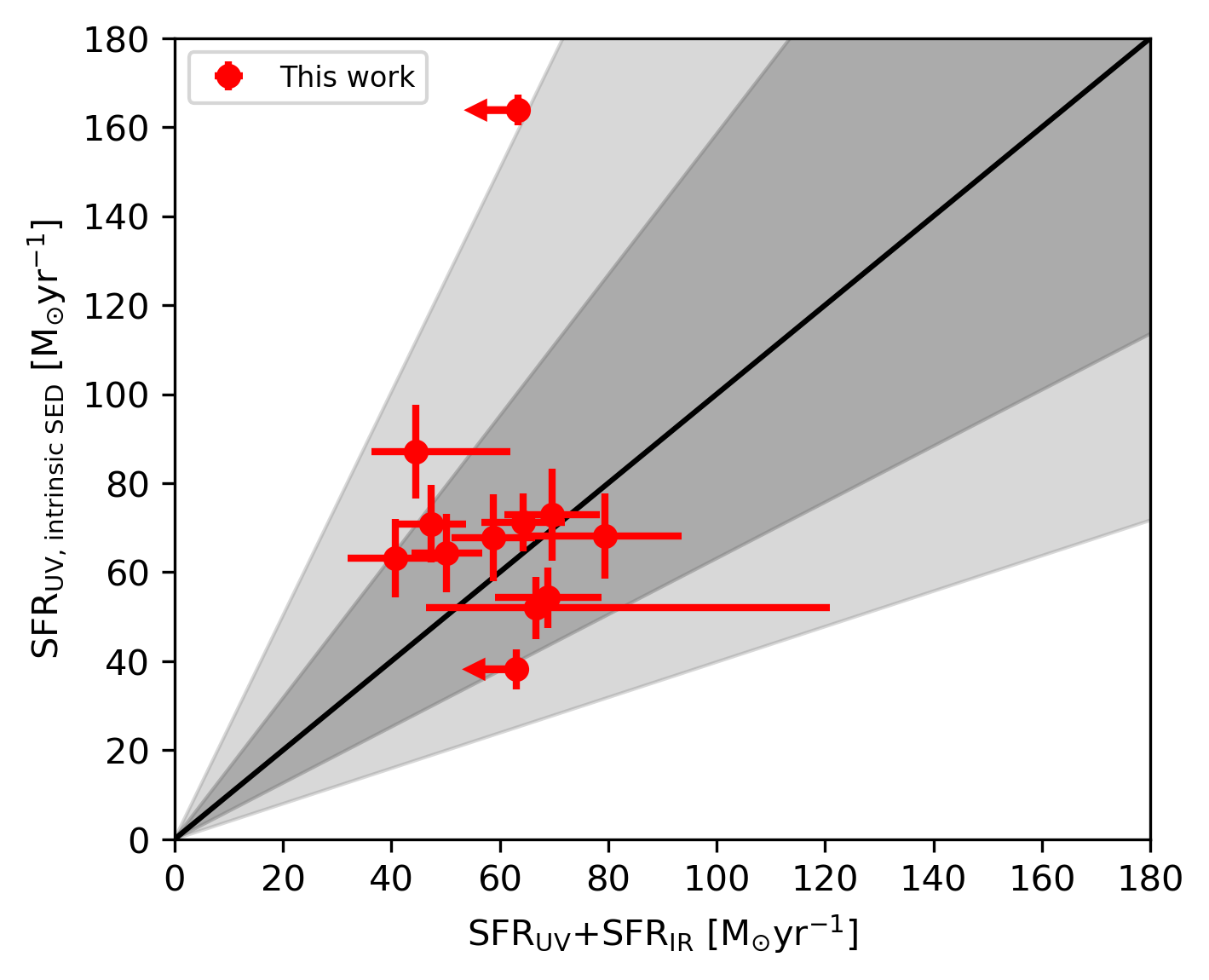}
    \caption{A comparison of two approaches for correcting rest-UV SFRs for dust attenuation.  
    In the left panel, we show the total SFRs derived using the best-fit {\irxb} relation shown in Fig.~\ref{fig:IRX}, SFR$_{\text{UV}}$+SFR$_{\text{IR, IRX}}$,
    and in the right panel we show the rest-UV SFRs from the reconstructed intrinsic SED, SFR$_{\text{UV, intrinsic SED}}$. 
    Both methods recover the SFR$_{\text{UV}}+$SFR$_{\text{IR}}$ values well, with the shaded regions denoting a $0.2$~dex and $0.4$~dex offset from the 1-to-1 relation.
    In the left panel, the scatter around the 1-to-1 line corresponds to the offsets of the points from the {\irxb} relation, as expected.  
    In the right panel, REBELS-15 is an outlier, suggesting dust-age degeneracies in the SED fitting have caused an overestimation of its dust content.
    }
    \label{fig:SFR_Comparisons1} 
\end{figure*}

\begin{table*}
	\centering
	\caption{The star formation rates derived from different tracers for the 12 REBELS-IFU galaxies at $z\simeq7$ studied in this work. 
    From left to right, these are the rest-UV SFR derived directly from the spectra with no dust correction, the dust-corrected rest-UV SFR derived from the reconstructed intrinsic SED, the dust-corrected H$\alpha$ SFR, the FIR SFR inferred from {\Lir}, the FIR SFR inferred using the best-fit {\irxb} relation in Fig.~\ref{fig:IRX}, the 100 and 10~Myr SFRs obtained from the SED fits with a non-parametric SFH, and the {\cii}SFRs calculated using the calibration of \citet{DeLooze2014} using the {\cii}luminosities presented in \citet{Algera2025} and Schouws et al. (in prep.).  See Section~\ref{sec:results_SFR_tracers} for a full description of each tracer.
    }
	\label{tab:pt2}
	\begin{tabular}[]{cccccccccr} 
        \hline
		ID & SFR$_{\text{UV}}$ & SFR$_{\text{UV, intrinsic SED}}$ & SFR$_{\text{H}\alpha}$ & SFR$_{\text{IR}}$ & SFR$_{\text{IR, IRX}}$ & SFR$_{100~\text{Myr}}$ & SFR$_{10~\text{Myr}}$ & SFR$_{\text{[CII]}~158~\mu\text{m}}$ &\\[2ex]
         & {\Msun}/yr & {\Msun}/yr & {\Msun}/yr & {\Msun}/yr & {\Msun}/yr & {\Msun}/yr & {\Msun}/yr & {\Msun}/yr & \\
        \hline
        \hline
        REBELS-05 & $12.4 \pm 0.8$ & $71 \pm 9$ & $113 \pm 31$ & $35\substack{+7 \\ -7}$ & $44 \pm 9$ & $25\substack{+2 \\ -2}$ & $111\substack{+15 \\ -15}$ & $60 \pm 8$ & \\[2ex]
        REBELS-08 & $17.7 \pm 0.5$ & $54 \pm 7$ & $120 \pm 44$ & $51\substack{+10 \\ -10}$ & $22 \pm 4$ & $15\substack{+2 \\ -2}$ & $90\substack{+11 \\ -12}$ & $65 \pm 10$ & \\[2ex]
        REBELS-12 & $28.2 \pm 0.7$ & $68 \pm 10$ & -- & $51\substack{+14 \\ -14}$ & $60 \pm 10$ & $28\substack{+4 \\ -4}$ & $107\substack{+24 \\ -19}$ & $89 \pm 36$ & \\[2ex]
        REBELS-14 & $26.1 \pm 0.9$ & $68 \pm 10$ & -- & $33\substack{+8 \\ -8}$ & $49 \pm 9$ & $19\substack{+3 \\ -2}$ & $103\substack{+23 \\ -16}$ & $32 \pm 9$ & \\[2ex]
        REBELS-15 & $28.5 \pm 0.9$ & $164 \pm 3$ & $185 \pm 50$ & $<35$ & $27 \pm 5$ & $25\substack{+2 \\ -2}$ & $248\substack{+18 \\ -16}$ & $17 \pm 4$ & \\[2ex]
        REBELS-18 & $21.8 \pm 0.5$ & $71 \pm 7$ & -- & $43\substack{+8 \\ -8}$ & $60 \pm 11$ & $37\substack{+4 \\ -3}$ & $109\substack{+15 \\ -16}$ & $93 \pm 7$ & \\[2ex]
        REBELS-25 & $12.0 \pm 0.5$ & $52 \pm 7$ & -- & $55\substack{+54 \\ -20}$ & $30 \pm 7$ & $11\substack{+2 \\ -2}$ & $90\substack{+18 \\ -17}$ & $138 \pm 10$ & \\[2ex]
        REBELS-29 & $19.7 \pm 0.7$ & $64 \pm 9$ & $57 \pm 18$ & $31\substack{+7 \\ -7}$ & $26 \pm 5$ & $31\substack{+6 \\ -4}$ & $91\substack{+14 \\ -11}$ & $48 \pm 8$ & \\[2ex]
        REBELS-32 & $9.1 \pm 0.7$ & $63 \pm 9$ & $92 \pm 25$ & $32\substack{+9 \\ -9}$ & $37 \pm 9$ & $24\substack{+4 \\ -4}$ & $89\substack{+20 \\ -14}$ & $68 \pm 7$ & \\[2ex]
        REBELS-34 & $24.9 \pm 0.5$ & $38 \pm 4$ & $57 \pm 33$ & $<38$ & $10 \pm 2$ & $22\substack{+2 \\ -3}$ & $52\substack{+8 \\ -7}$ & $60 \pm 17$ & \\[2ex]
        REBELS-38 & $19.6 \pm 0.9$ & $87 \pm 11$ & $161 \pm 68$ & $25\substack{+17 \\ -8}$ & $47 \pm 10$ & $42\substack{+8 \\ -6}$ & $132\substack{+25 \\ -22}$ & $148 \pm 14$ & \\[2ex]
        REBELS-39 & $28.2 \pm 1.0$ & $73 \pm 10$ & $140 \pm 36$ & $41\substack{+9 \\ -9}$ & $21 \pm 4$ & $22\substack{+2 \\ -2}$ & $110\substack{+13 \\ -10}$ & $69 \pm 22$ & \\
        \hline
	\end{tabular}
\end{table*}

Given the importance of dust-obscured star formation in our sources,
in Fig.~\ref{fig:SFR_Comparisons1} we test how well SFR$_{\text{UV+IR}}$ is recovered by other rest-UV methods.
This is also relevant for even higher redshift galaxies, where observations are typically limited to the rest-UV.  
In the left panel of Fig.~\ref{fig:SFR_Comparisons1}, we compare the SFR$_{\text{UV+IR}}$ values to those obtained using the {\irxb} relation fitted in Fig.~\ref{fig:IRX} to correct the unobscured rest-UV SFR.  
The values show good agreement with SFR$_{\text{UV+IR}}$, with the scatter around the 1-to-1 relation corresponding to the offsets of the points from the {\irxb} curve.  
This suggests that using the {\irxb} relation can recover the total SFRs well, although with a strong dependence on the accuracy of the intrinsic UV-slope and attenuation curve slope assumptions.

In the right panel of Fig.~\ref{fig:SFR_Comparisons1}, we compare the SFR$_{\text{UV, intrinsic SED}}$ values to SFR$_{\text{UV+IR}}$.
The SFRs derived from the reconstructed dust-free SED models (SFR$_{\text{UV, intrinsic SED}}$) also recover the total SFRs derived from the rest-UV and FIR fluxes (SFR$_{\text{UV+IR}}$) well, with a standard deviation around the 1-to-1 relation of 17.6~{\Msun}/yr compared to 17.8~{\Msun}/yr for SFR$_{\text{UV}}$+SFR$_{\text{IR, IRX}}$.  We note that the correlations are actually statistically weak in both cases, although this could be a result of the narrow dynamic range of SFRs we probe.
However, REBELS-15 is a strong outlier when using the intrinsic SED method.
The FIR dust continuum of REBELS-15 is undetected.  Assuming REBELS-15 has a similar dust temperature to the other sources, this suggests that the SED fitting is overestimating the dust attenuation in this source, possibly due to age-dust degeneracies.
This demonstrates the limitations of only using the rest-UV/optical emission of a galaxy and the value of multi-wavelength observations, even non-detections.
However, we note the possibility that the dust temperature of REBELS-15 could be higher, which would naturally explain the ALMA Band 6 non-detection \citep[e.g.][]{Bakx2020} and simultaneously increase its SFR$_{\text{IR}}$.
Another effect we must consider that could affect the recovery of the total SFRs is the impact of dust-star geometry.  
We find that the deviation of the points from the 1-to-1 line in the right panel of Fig.~\ref{fig:SFR_Comparisons1} correlates with molecular index, $I_m$ \citep[see][]{Ferrara2022}, where higher $I_m$ values are indicative of the rest-UV and FIR emitting regions being spatially decoupled. This is also seen in ALPINE galaxies \citep{Sommovigo2022b} and is suggestive of a non-uniform ISM morphology.  Future spatially-resolved analysis will investigate the effects of these spatial offsets further and their implications for the assumption of energy balance in the SED fitting (Fisher et al. in prep.).

The majority of the SFR$_{\text{[CII]}~158~\mu\text{m}}$ values show good consistency with the total SFR estimated from SFR$_{\text{UV}}+$SFR$_{\text{IR}}$.
However, we do not discuss the SFRs derived from {\cii}further due to the significant uncertainties on the \cite{DeLooze2014} calibration.

\subsection{The star-forming main-sequence with SFR$_{\text{UV+IR}}$}
\label{sec:discussion_MS}

\begin{figure} 
\includegraphics[width=\columnwidth]{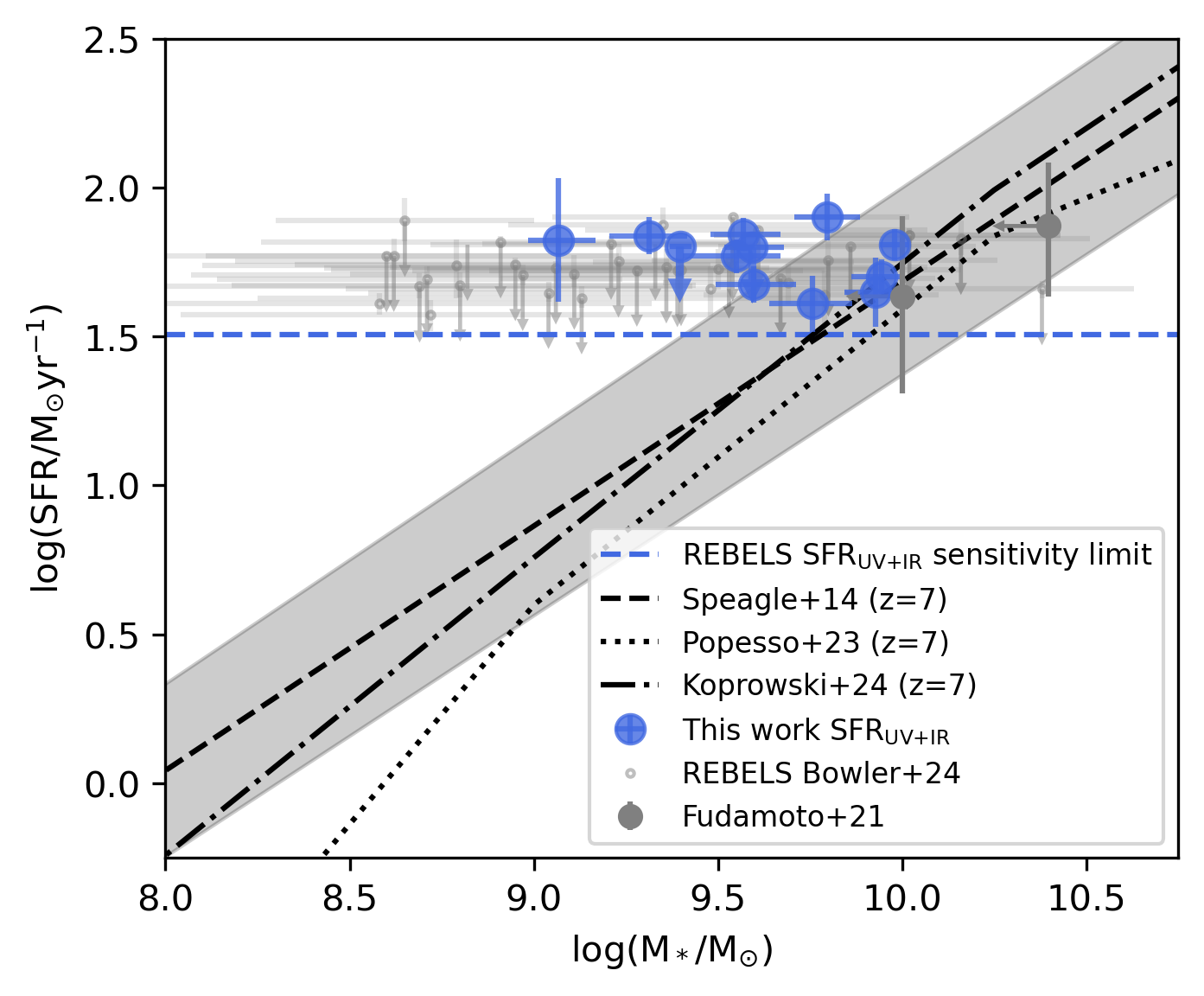}
\caption{
The star-forming main-sequence relation with the 12 massive, $z\simeq7$ REBELS-IFU galaxies is shown in blue.
The SFRs are derived by combining the unobscured SFR from the rest-UV and the obscured SFR from the ALMA FIR dust continuum observations.
The blue horizontal dashed line shows the SFR threshold of the REBELS survey imposed by the rest-UV bright selection and the sensitivity of the ALMA continuum observations.
The small grey points show the other REBELS galaxies \citep{Bowler2023} and the larger grey points show the serendipitous sources discovered in the REBELS ALMA data \citep{Fudamoto2021}.
The REBELS-IFU galaxies lie above the main-sequence relations from \citet{Speagle2014}, \citet{Popesso2022}, and \citet{Koprowski2024} shown by the dashed, dotted, and dash-dot lines, respectively.  These relations have been extrapolated to $z=7$ and are derived by combining a range of different tracers.}
\label{fig:SFR_MS_Plot1} 
\end{figure}

\begin{figure} 
\includegraphics[width=\columnwidth]{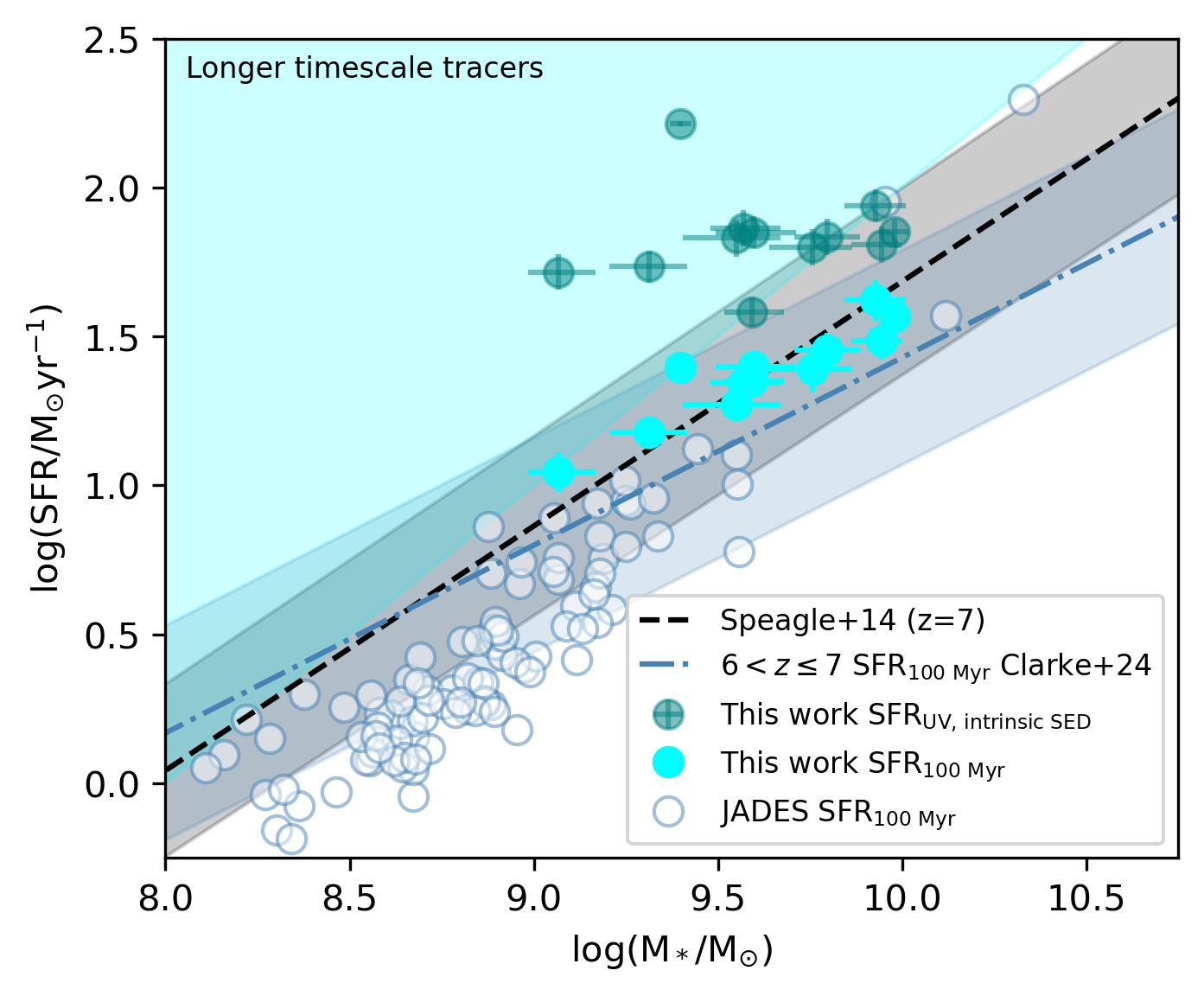}
\includegraphics[width=\columnwidth]{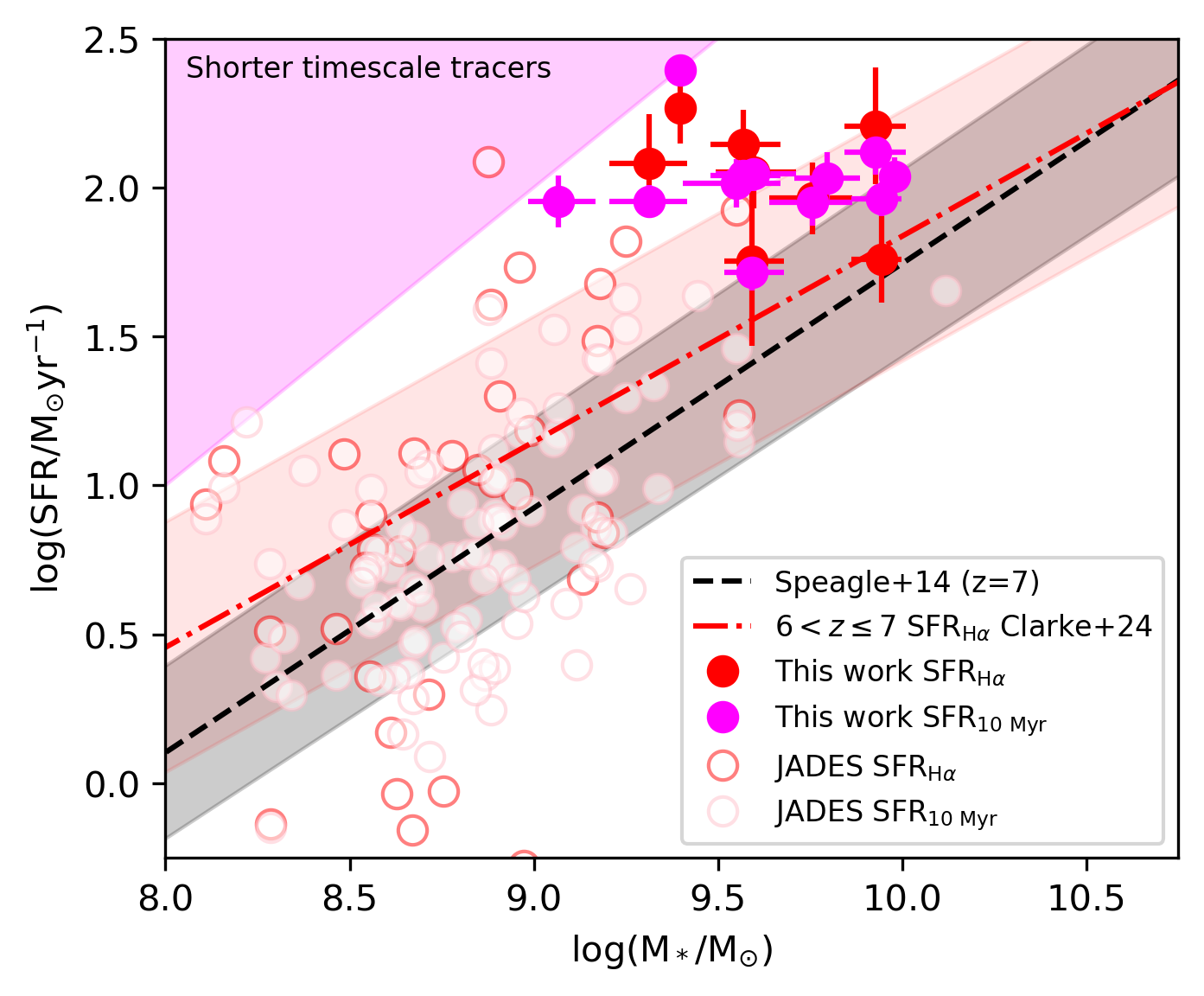}
\caption{
The star-forming main-sequence relation with longer (upper panel) and shorter (lower panel) timescale tracers.  
The cyan and magenta shaded areas show the forbidden regions of the parameter space for the $100$ and $10$~Myr timescales, derived by assuming all the stellar mass has been formed over that timescale \citep[][]{Carnall2018}.
In the upper panel, we show that the SFRs derived from the rest-UV flux of the reconstructed intrinsic SEDs, SFR$_{\text{UV, intrinsic SED}}$, in teal and the SFR$_{100~\text{Myr}}$ values (cyan) from the same SED fits. 
The SFR$_{\text{UV, intrinsic SED}}$ values systematically exceed the SFR$_{100~\text{Myr}}$ values. This discrepancy can be explained by rising SFHs (see Section~\ref{sec:discussion_SFH_model}).
In the bottom panel, we show the SFR$_{\text{H}\alpha}$ values in red and the SFR$_{10~\text{Myr}}$ values from SED fitting in magenta.
These two tracers agree well.  
The H$\alpha$ and $100$~Myr main-sequence relations from CEERS and JADES galaxies at $6<z\leq7$ are shown by the red and blue dash-dot lines \citep{Clarke2024}, and JADES galaxies in the same redshift range as the REBELS-IFU sample are shown by open circles. 
}
\label{fig:SFR_MS_Plot} 
\end{figure}

In Fig.~\ref{fig:SFR_MS_Plot1}, we show the total SFRs of the REBELS-IFU galaxies derived by combining the rest-UV and FIR tracers, SFR$_{\text{UV+IR}}$, in comparison to literature star-forming main-sequence relations.
SFR$_{\text{UV+IR}}$ is the most direct way of measuring the total SFR and is independent of dust attenuation curve assumptions.
However, it still depends on the assumed luminosity-to-SFR conversion factors and dust temperature. 
We note that even if we adopt the dust temperatures for these sources presented in \cite{Ferrara2022} or \cite{Sommovigo2022}, which vary between galaxies, the galaxies still lie systematically above the literature main sequence relations, and our key results are unaffected.
We also caveat that, as discussed at length in \cite{Sommovigo2025b}, assuming a single-temperature modified blackbody for the FIR dust SED can bias {\Lir} measurements.  
Better sampling of the FIR dust SEDs via multi-band ALMA observations is required to place stronger constraints on {\Lir} and thus the dust-obscured SFR.  
Recently obtained ALMA observations now provide dust continuum detections in two bands for the dust-detected galaxies in this sample (Algera et al. in prep.).
Preliminary analysis of these data suggests that we are adopting an appropriate dust temperature and that the SFR$_{\text{IR}}$ values remain consistent within the errors, meaning our findings will not significantly change.

The REBELS-IFU galaxies tend to lie slightly above the star-forming main sequence relations derived by \cite{Speagle2014}, \cite{Popesso2022}, and \cite{Koprowski2024} that have been extrapolated to $z=7$ (these relations were derived from galaxies up to $z\simeq6$) and adjusted to match the luminosity-to-SFR conversion factors we use.
We also show the other REBELS sources \citep{Bowler2023} and the serendipitous sources discovered in the ALMA data from the REBELS large program by \cite{Fudamoto2021}. 
The dashed horizontal line shows the approximate SFR threshold of the REBELS survey introduced by the rest-UV bright selection ({\muv}$\lesssim-21.5$) and the dust continuum sensitivity limits of the ALMA observations. 
This introduces a bias towards higher SFRs and an artificial flatness in the data points \citep[see also][]{Bouwens2022, Algera2022}. 
The 12 REBELS-IFU sources have some of the brightest {\cii}$158~\mu$m detections from the REBELS survey, further biasing this subsample to higher SFRs \citep[e.g.][]{DeLooze2014}.
If we simulate the main sequence using the relation and scatter from \cite{Speagle2014} and apply the selection cuts, the simulated galaxies occupy a similar region of the parameter space, as indicated by the horizontal line in Fig.~\ref{fig:SFR_MS_Plot1}.

\subsection{The star-forming main-sequence with other tracers}
\label{sec:discussion_MS_other_tracers}

In Fig.~\ref{fig:SFR_MS_Plot}, we show the positions of the REBELS-IFU sample with respect to the star-forming main-sequence using other SFR tracers, with tracers that are sensitive to longer (shorter) timescales in the top (bottom) panel.  
We plot the SFRs derived from the NIRSpec spectra of JADES galaxies in the same redshift range with open circles \citep{Eisenstein2023, Bunker2024, DEugenio2025}.
The REBELS-IFU galaxies tend to lie above all the main sequence relations derived using different tracers from CEERS and JADES spectra at $6~<~z~\leq~7$ by \cite{Clarke2024}.
The forbidden regions set by the maximum SFR averaged over a given timescale as a function of stellar mass that has been formed \citep[see][]{Carnall2018} are shown by the shaded regions.
For a galaxy to lie in these regions, it would require the SFR averaged over a given timescale to be greater than the total mass formed divided by the total elapsed time.

First we consider the SFR averaged over longer ($100$~Myr) timescales, comparing SFR$_{\text{UV, intrinsic SED}}$ and SFR$_{100~\text{Myr}}$ (upper panel). 
While SFR$_{\text{UV, intrinsic SED}}$ shows a flat relation similar to SFR$_{\text{UV+IR}}$, as expected given the relatively good agreement we found between these quantities in Fig.~\ref{fig:SFR_Comparisons1}, SFR$_{100~\text{Myr}}$ correlates strongly with stellar mass and shows extremely good consistency with the \cite{Speagle2014} relation and $100$~Myr SED-based relation from \cite{Clarke2024}.
As will be discussed further in Section~\ref{sec:discussion_SFH_model}, SFR$_{\text{UV, intrinsic SED}}$ and SFR$_{\text{UV+IR}}$ are calculated using $\kappa$ values that are derived assuming a constant SFH.
This results in these tracing the SFR averaged over timescales shorter than $100$~Myr, according to the rising SFHs of the SED fits.
Indeed, the majority of the SFR$_{\text{UV+IR}}$ and SFR$_{\text{UV, intrinsic SED}}$ values lie in the forbidden region of Fig.~\ref{fig:SFR_MS_Plot} and thus cannot be representative of an average star formation rate over $100$~Myr for these SED-derived stellar masses.

Second we consider the SFR averaged over shorter ($10$~Myr) timescales in the bottom panel, we find good agreement between the SFR$_{10~\text{Myr}}$ and SFR$_{\text{H}\alpha}$ values for the eight galaxies in the REBELS-IFU sample at $6.5\leq z < 7.0$ for which H$\alpha$ lies within the NIRSpec wavelength coverage.
As will be discussed further in Section~\ref{sec:discussion_SFH_model}, $\kappa_{\text{H}\alpha}$ changes less dramatically than $\kappa_{\text{UV}}$ for a rising SFH compared to a constant SFH and thus these continue to trace similar timescales.
On average, SFR$_{\text{H}\alpha}$ tends to slightly exceed SFR$_{\text{UV+IR}}$, consistent with the rising SFH seen in the SED fits, even though SFR$_{\text{UV+IR}}$ and  SFR$_{\text{UV, intrinsic SED}}$ are effectively tracing shorter timescales.
Similarly, the \cite{Clarke2024} main sequence relation derived using H$\alpha$ has a higher normalisation and larger scatter than the rest-UV derived relation.  
The range of SFR$_{\text{H}\alpha}$ values in our sample is also slightly larger than SFR$_{\text{UV+IR}}$, but our small sample size makes it challenging to comment on the scatter of the star-forming main-sequence and the implications of this for SFH burstiness.

\section{Discussion}
\label{sec:discussion} 
We have derived the SFRs from different tracers for 12 Lyman-break galaxies at $z = 6.5-7.7$ using \textit{JWST} NIRSpec spectra and ALMA observations, showing that the majority of these tracers place the REBELS-IFU galaxies above literature $z=7$ star-forming main-sequence relations.
These galaxies have high dust-obscured SFR fractions, but our multi-wavelength data allow us to robustly correct for this.  
We will discuss the physical reasons for discrepancies between SFR tracers and show that the REBELS-IFU SFHs rise more steeply than lower mass galaxies at the same redshift.
We illustrate the effects of these rising SFHs on luminosity-to-SFR conversion factors using SED models and present new conversions more relevant for $z\simeq7$ galaxies.
Finally, we use these results to show that the H$\alpha$-to-UV luminosity ratio is an unreliable probe of SFH burstiness.

\subsection{Tracer timescales and rising SFHs cause SFR discrepancies}
\label{sec:discussion_timescales}

\begin{figure} 
    \includegraphics[width=\columnwidth]{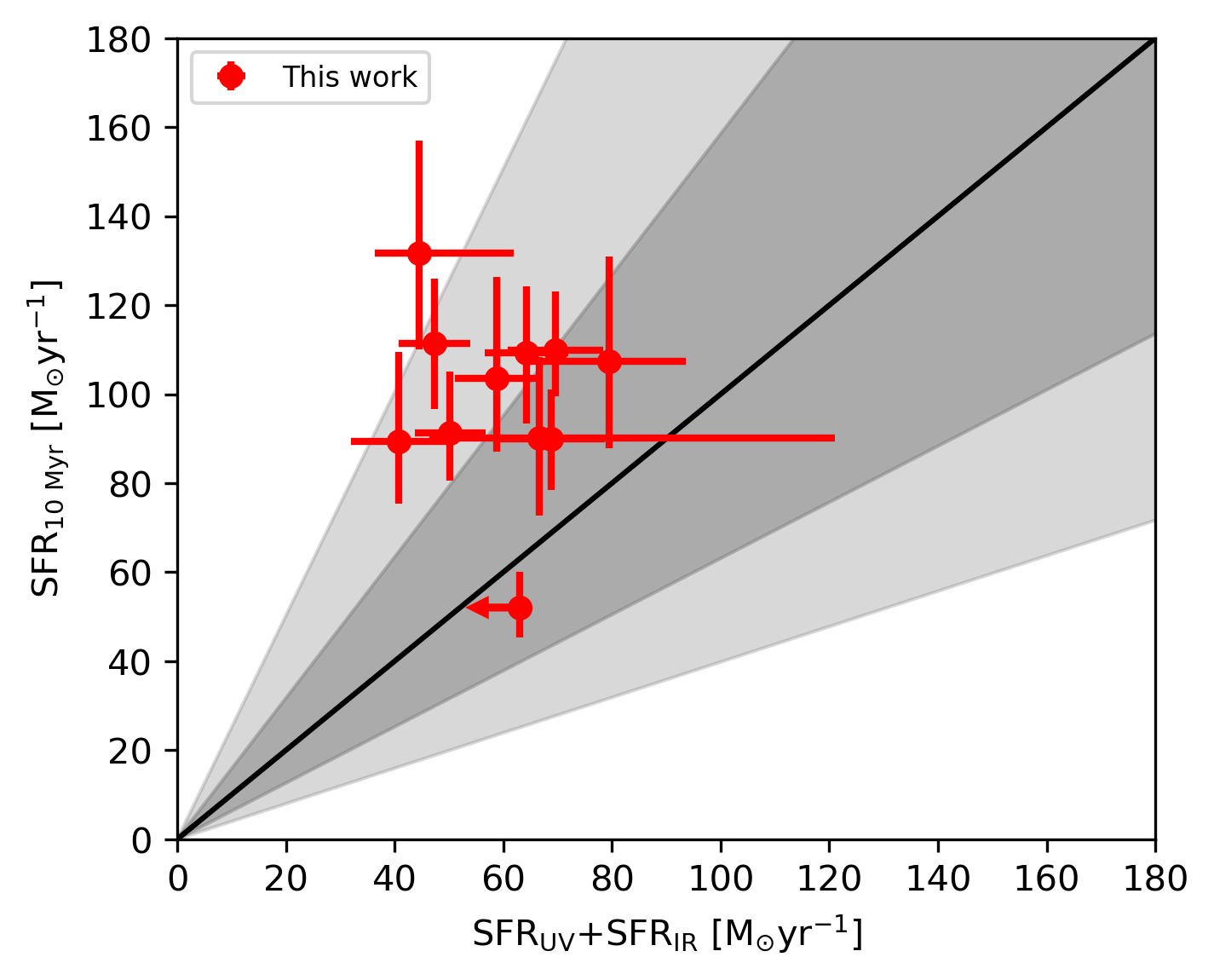}
    \includegraphics[width=\columnwidth]{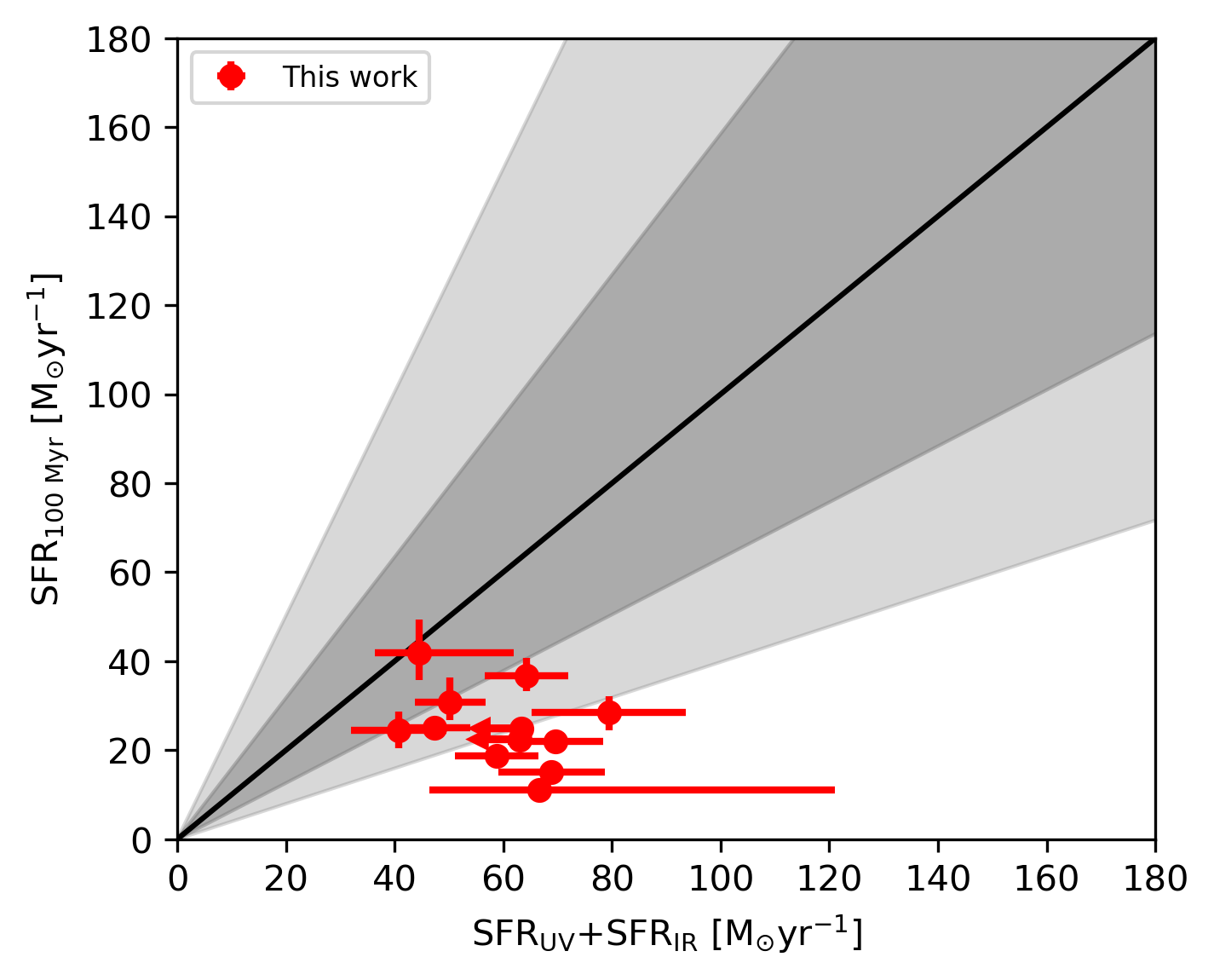}
    \caption{
    We show that the SFR$_{\text{UV+IR}}$ values for the REBELS-IFU galaxies are systematically offset below the SED-derived $10$~Myr SFRs (top panel) and above the $100$~Myr SFRs (bottom panel).
    This implies the SFR$_{\text{UV+IR}}$ values, calculated using luminosity-to-SFR conversion factors that are derived assuming a constant SFH, are tracing the average SFR over a timescale between these two values. 
    According to the non-parametric SFHs of our SED fits, our SFR$_{\text{UV+IR}}$ values trace the SFR averaged over a median timescale of $\sim20$~Myr.
    }
    \label{fig:SFR_Comparisons} 
\end{figure}

In Fig.~\ref{fig:SFR_MS_Plot}, we showed the discrepancy between SFR$_{\text{UV, intrinsic SED}}$, derived using the luminosity-to-SFR conversion factor given in Table~\ref{tab:SFREqns}, and the SFR$_{100~\text{Myr}}$ values derived from the same SED fits.  
The offset between these is greater for the galaxies with more steeply rising SFHs (i.e. it anticorrelates with the mass-weighted ages of the SED fits, such that the discrepancy is larger for younger ages with lower mass-to-light ratios), as will be discussed in more detail in Section~\ref{sec:discussion_SFH_model}
In Fig.~\ref{fig:SFR_Comparisons} we show that the inferred $10$~Myr averaged SFRs from the SED fitting are systematically higher than SFR$_{\text{UV+IR}}$, consistent with a rising SFH and SFR$_{\text{UV+IR}}$ probing timescales longer than $10$~Myr.
However, the $100$~Myr SFR values from the SED fitting are systematically lower than the SFR$_{\text{UV+IR}}$ values, which is puzzling as SFR$_{\text{UV+IR}}$ is expected to probe this timescale.
This discrepancy is a result of using conversion factors that assume a constant SFH in the previous $100$~Myr.
As will be shown in more detail in Section~\ref{sec:discussion_SFH_model}, the $\kappa$ values needed to convert between, for example, SFR$_{100~\text{Myr}}$ and {\Luv} depend on the exact form of the SFH, especially in the rest-UV
\citep[see also Fig. 10 in][]{Pallottini2022}.

$100$~Myr SFRs that are lower than rest-UV derived SFRs have also been found in other observational works.
For example, \cite{Pirie2024} finds dust-corrected SFR$_{\text{UV}}$ values that are enhanced by a factor of $2-3$ compared to the SED-derived SFR$_{100~\text{Myr}}$ values derived from the NIRCam narrow-band imaging of the JELS (\textit{JWST} Emission Line Survey) galaxies at $z\simeq6.1$, with the largest offsets in galaxies with lower stellar masses and enhanced recent star formation.
We find comparable factors in the REBELS-IFU sample, with SFR$_{\text{UV+IR}}$ and SFR$_{\text{UV, intrinsic SED}}$ exceeding SFR$_{\text{100 Myr}}$ by median factors of 2.3 and 2.7, respectively.  
Thus, there is growing observational evidence that at these redshifts the rest-UV continuum light of galaxies is dominated by fluctuations in star formation on timescales shorter than $100$~Myr, introducing biases in SFRs derived from the rest-UV using $\kappa$ values that assume constant SFHs.
Simulations have also shown that stochastic variability in SFRs can cause rest-UV magnitude variations at these redshifts \citep[e.g.][]{Pallottini2023}.

These observational results are also consistent with the \textsc{THESAN-ZOOM} hydrodynamics simulations presented in \cite{McClymont2025}.
They find that the $100$~Myr SFRs are below the rest-UV SFRs derived using $\kappa$ values that assume a constant SFH, with the rest-UV SFRs instead probing a timescale of $24$~Myr.  
If we take the non-parametric SFHs of our SED fits, we find that SFR$_{\text{UV+IR}}$ is equivalent to the SFR averaged over a median timescale of $\sim20$~Myr.
Similarly, results from the FIRE simulations suggest that, since the integrated rest-UV light of a galaxy is dominated by the young, massive, and short-lived stars, the rest-UV continuum can trace timescales ranging between $10$~Myr and more than $100$~Myr, depending on the SFH \citep{FloresVelazquez2021}.
This suggests that variability in star formation on timescales shorter than $100$~Myr is introducing scatter and biases into SFRs inferred from the rest-UV emission, introducing scatter to main sequence relations since the timescales probed are dependent on the individual galaxy SFH.  
The impact of this is likely to increase at higher redshifts, given the evidence that SFHs are more likely to be bursty or rising \citep[e.g.][]{Tacchella2018, Cole2023, Carvajal-Bohorquez2025, Simmonds2025}, although the simulations of \cite{Pallottini2025} suggest SFH stochastity cannot be too high without destroying the observed mass metallicity relation.
Thus, SFRs derived from the rest-UV/FIR using calibrations that assume a constant SFH should not be assumed to be equivalent to the $100$~Myr average SED result in high-redshift galaxies since they are assuming different SFHs to convert between the observed flux and SFR.
The SED-derived SFRs are arguably more informative than the SFR$_{\text{UV}}$ or SFR$_{\text{H}\alpha}$ values derived in this work since they use all the information in the observed NIRSpec spectra and allow for some flexibility in the SFHs, compared to the very limited wavelength ranges and the (likely invalid) assumptions made to derive the luminosity-to-SFR conversion factors (e.g. constant SFH and fixed metallicity) used for the latter two. 

\begin{figure}
\includegraphics[width=\columnwidth]{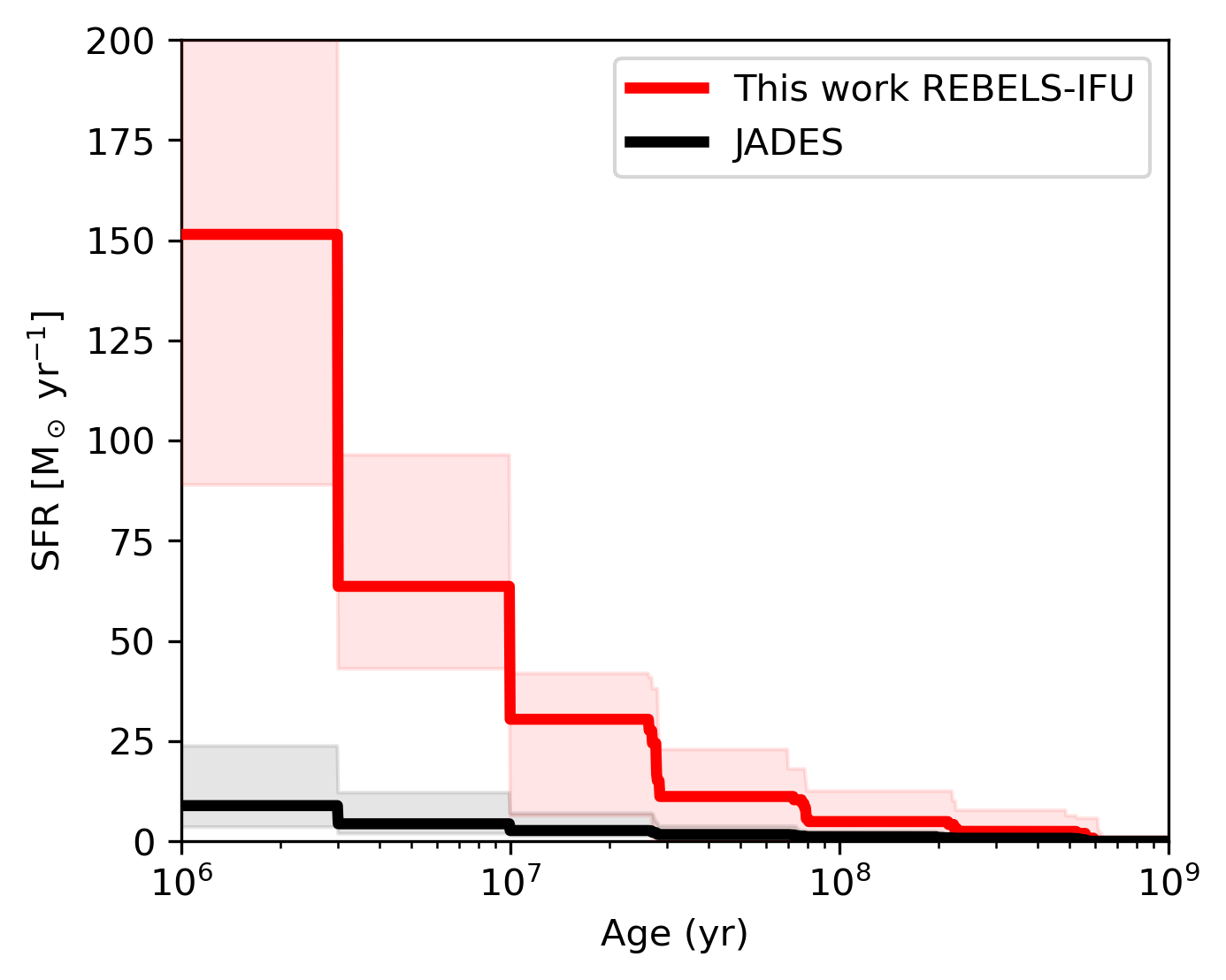}
\includegraphics[width=\columnwidth]{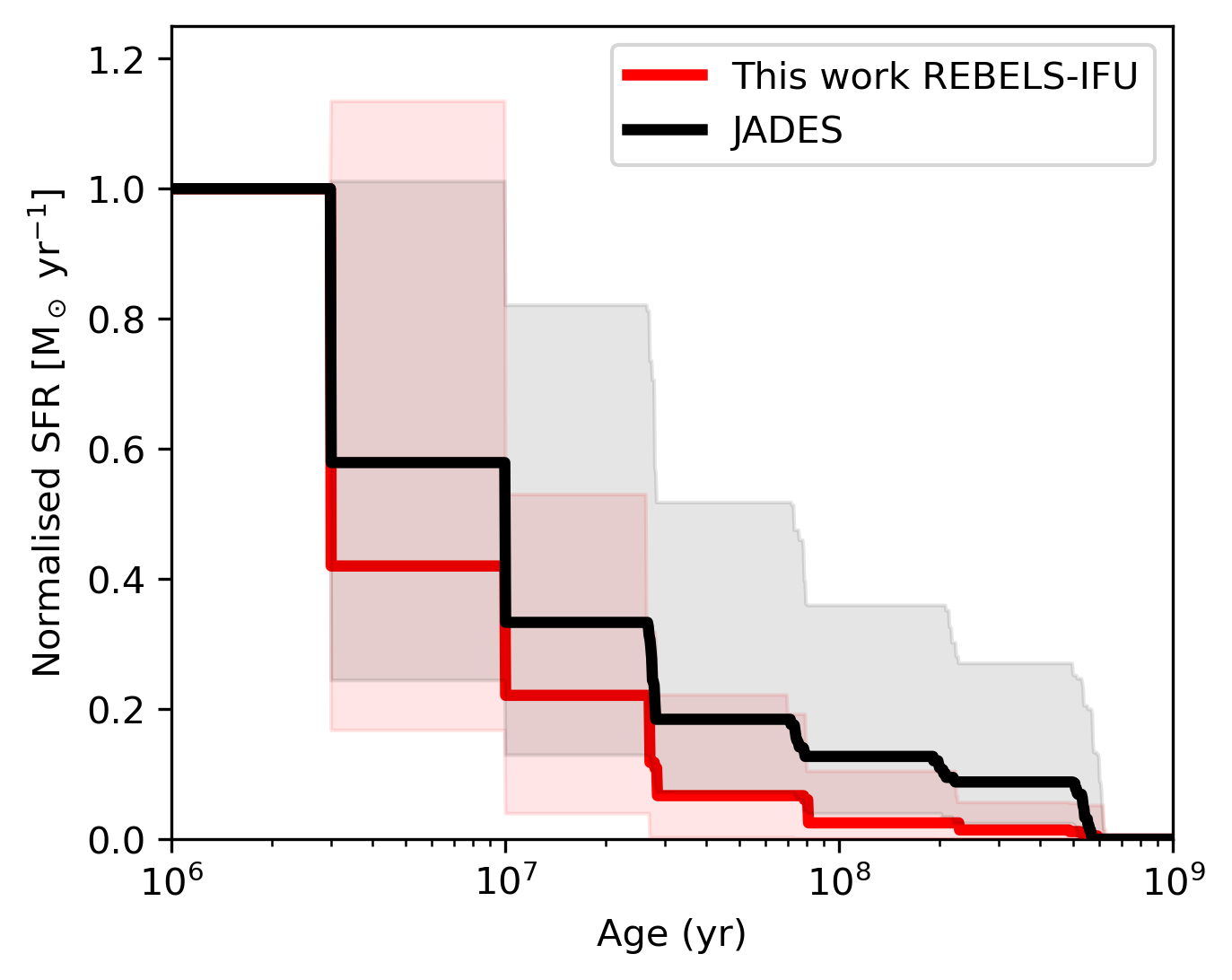}
\caption{
In the top panel, we show the median SFH of the REBELS-IFU galaxies in red.
In black, we show the median SFH of JADES galaxies in the same redshift range.   
The REBELS-IFU SFHs are significantly higher than the JADES galaxies.  
In the bottom panel, we normalise the SFHs by the most recent bin to compare their shapes and find that the REBELS-IFU SFHs tend to rise more steeply.  
The shaded regions show the 16th-84th percentile range.}
\label{fig:SFH_comparison_JADES} 
\end{figure}

\subsection{Star formation histories}
\label{sec:discussion_NP_SFHs}

To further understand the discrepancies between SFR tracers and the positions of the REBELS-IFU galaxies relative to the main sequence relations, we investigate their SFHs. 
SED fits with a constant SFH constrained to have a minimum timescale of at least $100$~Myr fit the spectra poorly.
In particular, the lack of a Balmer Break feature in the observed integrated spectra around rest-frame wavelengths of $4000$~{\AA}, which is associated with older stars, most notably A-type stars, causes the optical continuum level and emission lines to be poorly fitted \citep[e.g.][]{Bruzual1983, Poggianti1997}, suggesting the contribution of these stars to the integrated spectrum is not significant. 
Our findings are consistent with the work of \cite{Topping2022} on the REBELS galaxies, which also finds that SED fitting with a constant SFH results in younger ages and masses that are systematically lower than non-parametric fits. 
On average, our non-parametric masses are $0.35$~dex higher than those from the constant SFH, which is similar to the offset of $0.43$~dex \cite{Topping2022} finds.
We note that stellar masses derived from integrated spectra can be underestimated compared to pixel-based estimates due to outshining \citep[e.g.][]{Gimenez-Artega2024}. 
However, these effects are not expected to be significant in high-mass galaxies like these \citep[e.g.][]{Lines2024}.

In Fig.~\ref{fig:SFH_comparison_JADES}, we compare the average non-parametric SFH of the REBELS-IFU galaxies\footnote{We note that we present an average SFH for the REBELS-IFU sources using the integrated spectra. A detailed analysis of the SFH of each galaxy will be presented in Laza-Ramos et al. in prep., including spatially-resolved component-by-component analysis using the IFU data, which will provide a more comprehensive picture of how these galaxies have built up their stellar mass.} to the JADES galaxies in the same redshift range \footnote{We fit the NIRSpec spectra of the JADES galaxies with the same BAGPIPES setup as in \cite{Fisher2025}, except we fix the attenuation curve to \cite{Calzetti2000} since we expect their dust content to be lower and thus the attenuation curve slope recovery to be less reliable \citep[see Appendix A of ][]{Fisher2025}.}. 
We note that the stellar mass distributions of these two samples are not the same, with the median mass of these JADES galaxies being $\log(${\mstar}/{\Msun}$)=8.8$ compared to $\log(${\mstar}/{\Msun}$)=9.6$ for the REBELS-IFU sample. 
In the top panel of Fig.~\ref{fig:SFH_comparison_JADES}, we find, unsurprisingly, that the median SFH of the REBELS-IFU galaxies is significantly higher than the JADES galaxies.
In the bottom panel of Fig.~\ref{fig:SFH_comparison_JADES}, we plot the SFHs normalised by the most recent bin.  
Although the variation between the galaxies, indicated by the shaded regions, is quite large, the REBELS-IFU SFHs, on average, rise more steeply than the JADES galaxies.
Given that lower stellar mass galaxies are generally thought to have more steeply rising or stochastic SFHs \citep[e.g.][]{Looser2023, Legrand2021}, this is unexpected.
However, the bright rest-UV selection bias could go some way to explaining this, with work from the JADES survey by \cite{Endsley2024b} suggesting that the brightest galaxies in their sample are often experiencing a recent strong upturn in their SFRs. We also note that steeply rising SFHs are consistent with the results of the SERRA simulations presented in \cite{Kohandel2025}.

\begin{figure} 
\includegraphics[width=\columnwidth]{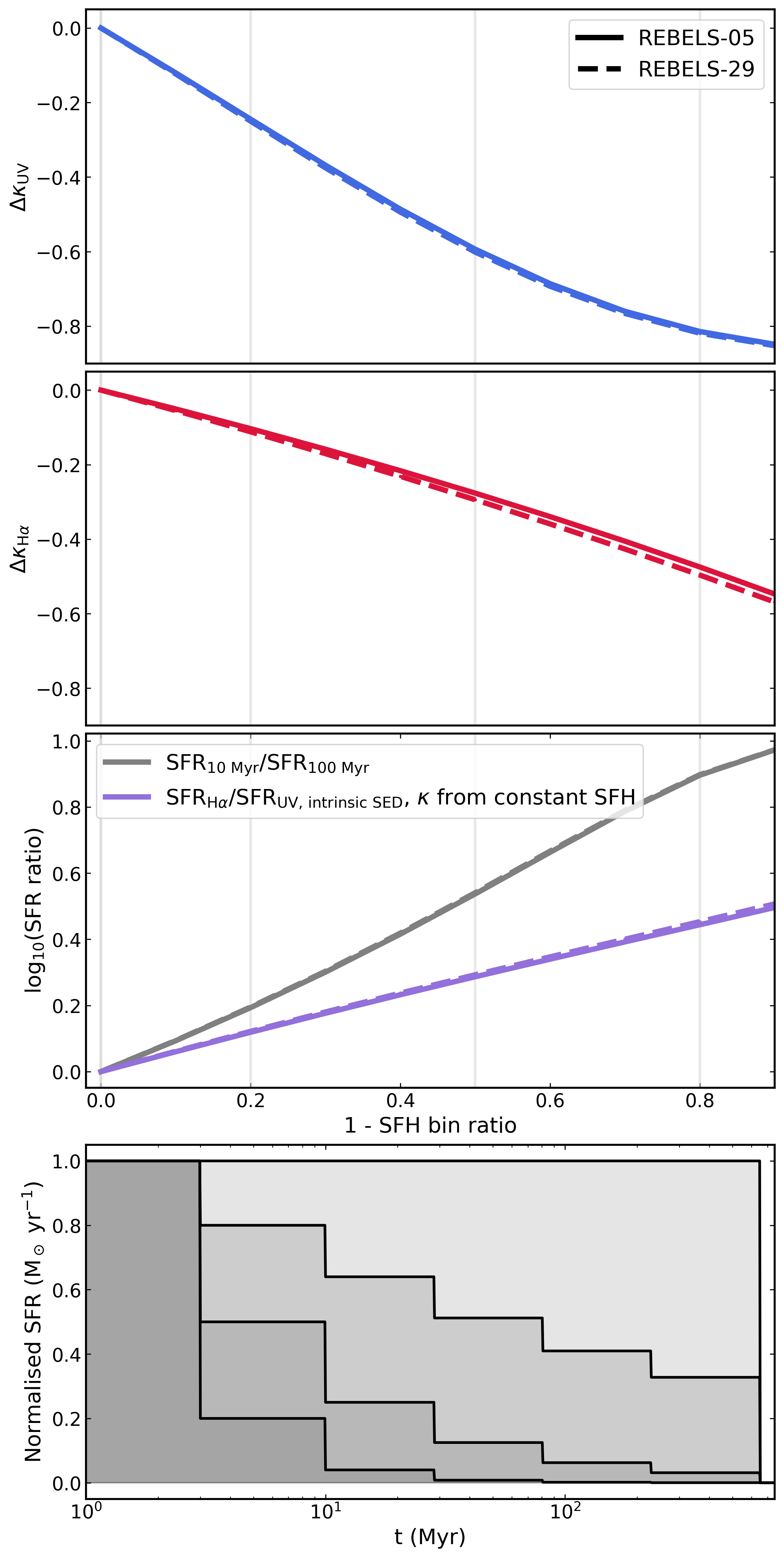}
\caption{
We create model galaxy spectra using BAGPIPES with non-parametric SFHs based on the SED fits to two of the REBELS-IFU galaxies. 
In the bottom panel, we show the SFHs of increasing steepness, normalised by the most recent SFH bin. 
The grey vertical lines in the top three panels mark the locations of the constant and three rising SFHs shown. 
In the top two panels, we show the fractional change in the luminosity-to-SFR conversion factors, $\kappa$ (calculated using  $\kappa_{\text{UV}} =$ SFR$_{100~\text{Myr}}/${\Luv} and  $\kappa_{\text{H}\alpha} =~$SFR$_{10~\text{Myr}}/L_{\text{H}\alpha}$) compared to the value obtained using a constant SFH against $1 - $ the SFR ratio between succesive time bins in the SFH, i.e. against increasing SFH steepness.
The conversion factors both decrease with increasing steepness of the rising SFH, with the variation in $\kappa_{\text{UV}}$ being more significant.
In the third row we show that the SFR$_{10~\text{Myr}}$/SFR$_{100~\text{Myr}}$ and 
SFR$_{\text{H}\alpha}$/SFR$_{\text{UV, intrinsic SED}}$ ratios, with the latter assuming the $\kappa$ values derived from the constant SFH model.
These both correlate with the steepness of the SFH, but the SFR$_{\text{H}\alpha}$/SFR$_{\text{UV, intrinsic SED}}$ ratio varies to a lesser degree since the effective timescales each tracer probes become more closely spaced in time as burstiness increases.}
\label{fig:SFH_model} 
\end{figure}

\subsection{Simulating the effect of rising SFHs on SFR tracers and $\kappa$}
\label{sec:discussion_SFH_model}

Given the strong evidence that our galaxies have non-constant SFRs in the $100$~Myr prior to observation, we further investigate the effect of a rising SFH on our results by generating model galaxy SEDs with BAGPIPES \citep{Carnall2018, Carnall2019}.
In Fig.~\ref{fig:SFH_model}, we show a model where we take the fitted SED of a REBELS-IFU galaxy and set the SFH such that the SFR in each time bin in the SFH differs from the previous by a constant ratio.  
We adjust this bin ratio to create SFHs of varying steepness, keeping all other parameters the same.
We also create a model galaxy that has a constant SFH over the same timescale.

Using the rest-UV and H$\alpha$ fluxes from the model SEDs, we calculate the $\kappa$ values required to convert between these luminosities and the known average SFR over $100$ or $10$~Myr (SFR$_{100~\text{Myr}}$ or SFR$_{10~\text{Myr}}$).  
These are shown in the top two panels.
We see $\kappa_{\text{UV}}$ and $\kappa_{\text{H}\alpha}$, which are the ratios between the SFRs and luminosities, both decrease with increasing SFH steepness.
This is to be expected because galaxies with more steeply rising SFHs have a greater fraction of their very young, massive stars formed in the more recent past, which dominate the spectra, resulting in greater luminosities for a given time-averaged SED SFR.  
The fractional decrease in $\kappa_{\text{UV}}$ occurrs more rapidly
\footnote{We note that in contrast to \cite{Pallottini2022}, our conversion factor decreases with increasing SFH steepness.  This is due to the fact that we calculate $\kappa_{\text{UV}}$ using SFR$_{\text{100 Myr}}$, whereas \cite{Pallottini2022} uses the SFR averaged over $20$~Myr.  Thus, in our case, the rate at which SFR$_{\text{100 Myr}}$ increases is slower than the rate at which $L_{\text{UV}}$ increases.}.

In the third row, we show how the SFR$_{\text{H}\alpha}$/SFR$_{\text{UV, intrinsic SED}}$ ratio and the SFR$_{10~\text{Myr}}$/SFR$_{100~\text{Myr}}$ ratio respond to increasing SFH steepness.
SFR$_{\text{H}\alpha}$ and SFR$_{\text{UV, intrinsic SED}}$ are calculated using the luminosity-to-SFR conversion factors ($\kappa$ values) inferred from the constant SFH, consistent with what is commonly assumed in calibrations.  
Both SFR ratios correlate with steepness, but the change in the SFR$_{\text{H}\alpha}$/SFR$_{\text{UV, intrinsic SED}}$ ratio is less pronounced than the SFR$_{10~\text{Myr}}$/SFR$_{100~\text{Myr}}$ ratio.
This is driven by the fact that $\kappa_{\text{UV}}$ varies more quickly than $\kappa_{\text{H}\alpha}$ and thus the timescales effectively probed when using $\kappa$ values from the constant SFH converge.  
We discuss further in Section~\ref{sec:discussion_burstiness} the implications of this for inferring whether a galaxy has a bursty SFH.

\subsection{Deriving new luminosity-to-SFR conversion factors}
\label{sec:discussion_new_kappas}
Given the non-constant nature of the REBELS-IFU SFHs, we derive new luminosity-to-SFR conversion factors from our rising non-parametric SFHs.  
To convert the observed rest-UV flux, corrected for dust attenuation, to the SFR averaged over $100$~Myr inferred from the SED fits (SFR$_{100~\text{Myr}}$) would require a median luminosity-to-SFR conversion factor of 
\begin{equation}
    \kappa_{\text{UV}}=(2.7\pm0.9) \times 10^{-29}~{\text{\Msun}}~\text{yr}^{-1}~\text{erg}^{-1}~\text{s Hz}
    \label{eq:SFR_UV_kappa}.
\end{equation}
To convert the dust-corrected H$\alpha$ flux to the SFR averaged over $10$~Myr inferred from the SED fits (SFR$_{10~\text{Myr}}$) would require a median luminosity-to-SFR conversion factor of 
\begin{equation}
\kappa_{\text{H}\alpha}= (4.7\pm1.4) \times 10^{-42}~{\text{\Msun}}~\text{yr}^{-1}~\text{erg}^{-1}~\text{s}
\label{eq:SFR_Ha_kappa}.
\end{equation}
The standard deviations in the $\kappa$ values are around $30$ per cent of the median values, which perhaps should be considered a lower limit on the uncertainty in $\kappa$ values such as these due to factors such as the variation in the shape of the galaxy SFHs and metallicity.   
Comparing these to the values in Table~\ref{tab:SFREqns} that are derived assuming constant SFHs, we see that $\kappa_{\text{H}\alpha}$ remains approximately consistent, but there is a large systematic shift in $\kappa_{\text{UV}}$ by a factor of $\simeq2.7$ ($>0.4$~dex). 
Thus, the SFRs derived from the rest-UV with the $\kappa$ values from a constant SFH would overestimate the $100$~Myr SFR by this factor.
This has significant implications for deriving SFRs for galaxies at even higher redshifts than our sample, since observations of these sources typically only probe the rest-UV and rising SFHs are increasingly likely \citep[e.g.][]{Simmonds2025}. 
By extension, $\kappa_{\text{IR}}$ will be affected in the same way as $\kappa_{\text{UV}}$.
Therefore, to avoid overestimating the SFR over a given timescale in galaxies with rising SFHs, adjusted conversion factors like these should be used.  
We note that the conversion factors applicable to the more abundant lower-mass galaxies at $z\simeq7$, which, as shown in Fig.~\ref{fig:SFH_comparison_JADES}, may have slightly less steeply rising SFHs, may not be quite as low, and therefore the conversion factors used should be informed by the SFH steepness (see Fig.~\ref{fig:SFH_model}).

\begin{figure*} 
\includegraphics[width=\columnwidth]{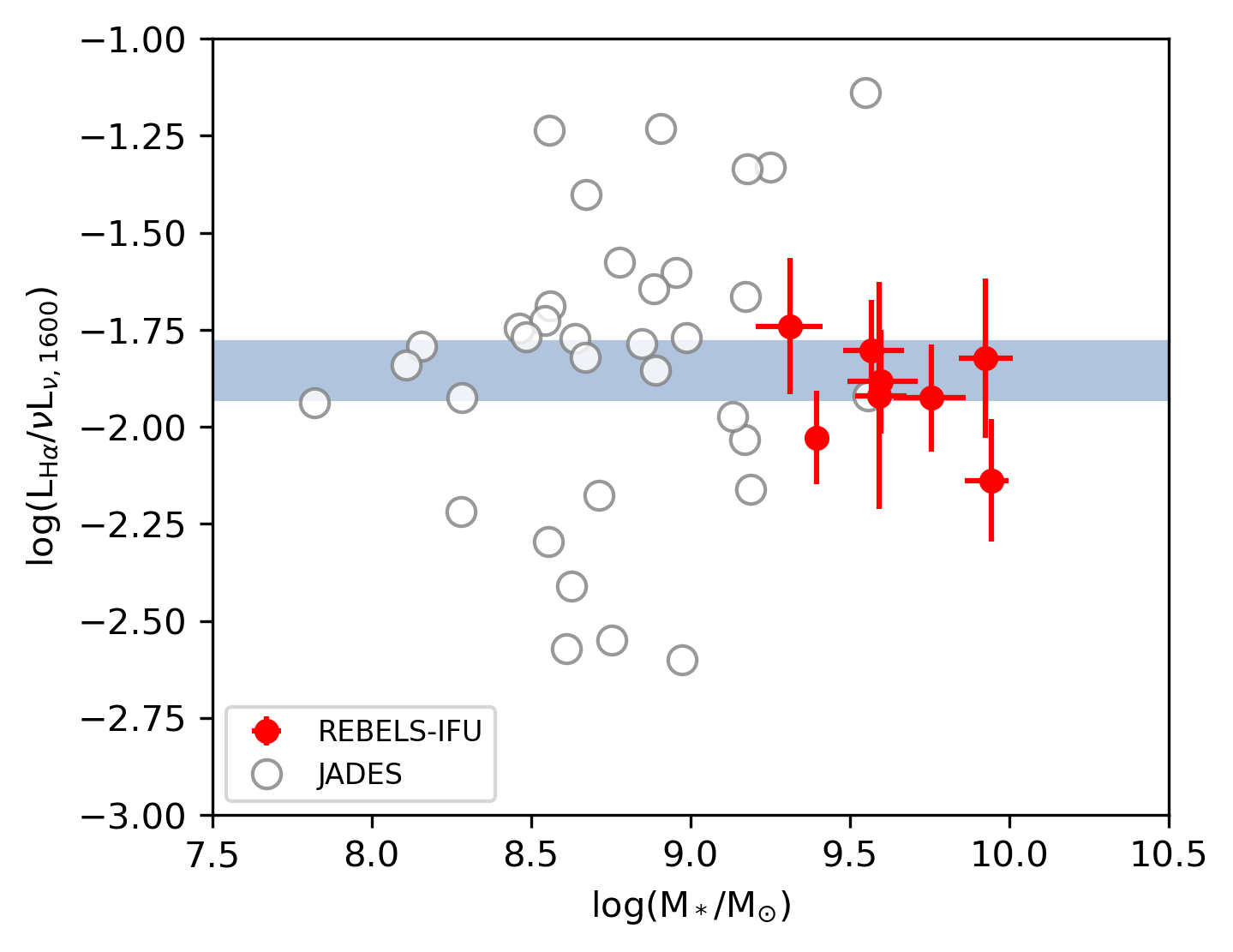}
\includegraphics[width=\columnwidth]{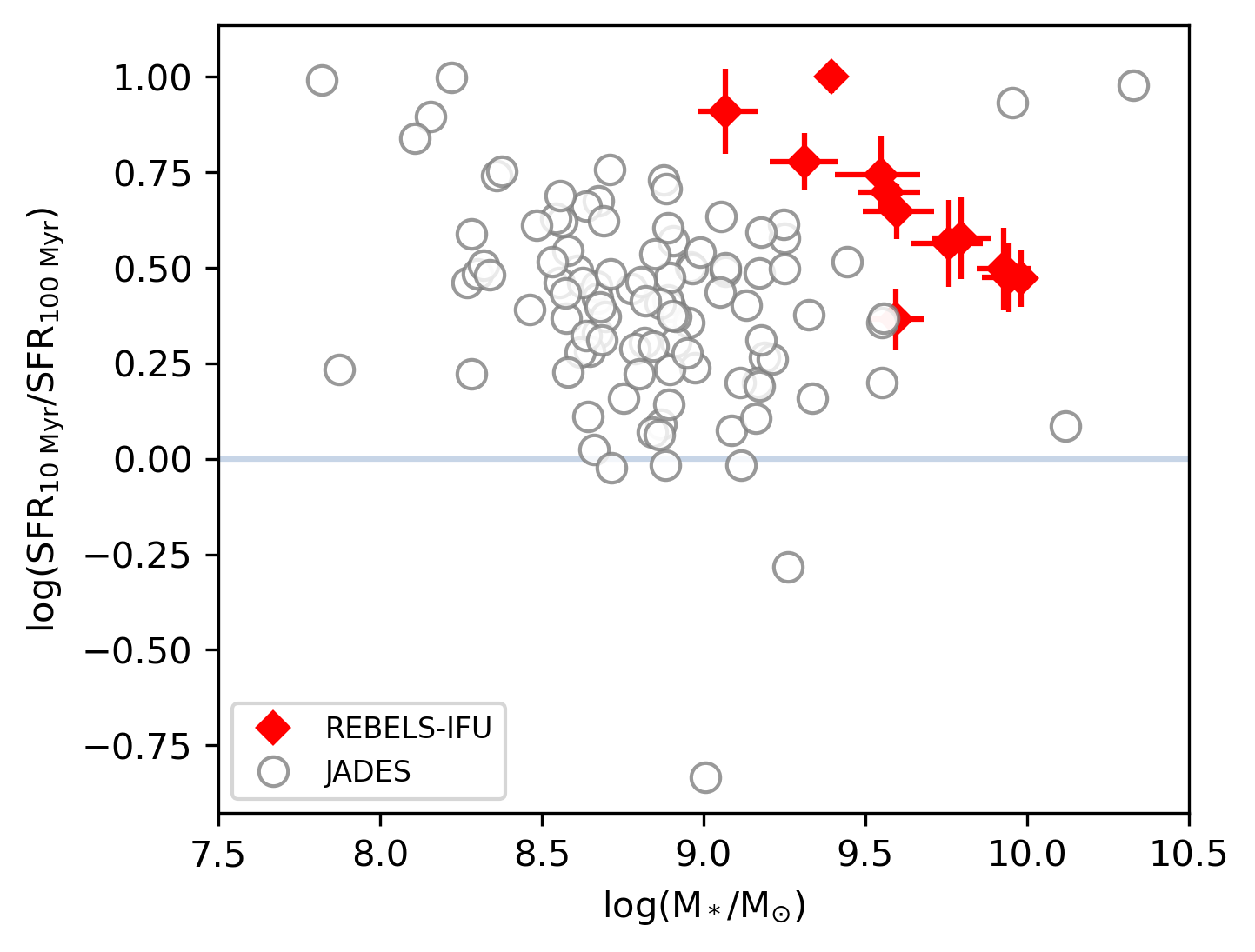}
\caption{
In the left panel, we plot the ratio of the dust-corrected H$\alpha$-to-UV luminosities as a function of stellar mass in red for the 8 of the 12 massive galaxies in the REBELS-IFU sample at $6.5\leq z < 7.0$ for which H$\alpha$ lies within the NIRSpec wavelength coverage.
The ratios are consistent with or below the equilibrium range from \citet{Mehta2023} shown by the blue shaded band, which indicates non-bursty SFHs. 
JADES galaxies at the same redshift as the REBELS-IFU sample are shown as open grey circles.  
In the right panel, we show that all the REBELS-IFU galaxies have $\log($SFR$_{10~\text{Myr}}$/SFR$_{100~\text{Myr}}) > 0$, indicative of rising or bursty SFHs, as derived from the SED fitting to the full spectra.
This suggests that the H$\alpha$-to-UV luminosity ratio can be consistent with non-bursty SFH values even when there is strong evidence for rising SFHs in the $100$~Myr prior to observation, implying it cannot reliably identify all bursty SFHs.  
}
\label{fig:burstiness} 
\end{figure*}

\subsection{How useful, therefore, are burstiness diagnostics?}
\label{sec:discussion_burstiness}

Finally, we use our results to assess the usefulness of the ratio of the H$\alpha$ luminosity, $L_{\text{H}\alpha}$, to the monochromatic luminosity at rest-frame wavelength 1600~{\AA}, $\nu L_{\nu, 1600}$, as a diagnostic for probing the SFHs of our galaxies \footnote{We note that this quantity is closely related to the ionising photon production efficiency, $\xi_{\text{ion, 0}}$, presented for this sample in Komarova et al. in prep.  The values presented there are consistent with what we derive in this work.}.
This ratio is commonly used in the literature in both local and high-redshift galaxies to quantify SFH burstiness \citep[e.g.][]{Meurer2009, Weisz2012, Guo2016, Emami2019, Faisst2019, Clarke2024} with some recent \textit{JWST} studies suggesting lower mass galaxies ($\log($\mstar/\Msun$)<9$) with higher emission line equivalent widths are more bursty \citep[e.g.][]{Atek2022, Pirie2024, Asada2023}.  
In Fig.~\ref{fig:burstiness}, we show the position of the REBELS-IFU galaxies using the dust-corrected rest-frame UV luminosities.
\footnote{We use the rest-UV flux derived from the intrinsic SED model over that derived from SFR$_{\text{UV+IR}}$/$\kappa_{\text{UV}}$ since the ratios using the latter are only available for 6 galaxies by the time we have excluded the four without H$\alpha$ coverage and the two without ALMA dust continuum detections.} 
We see that all the REBELS-IFU galaxies are consistent within 1$\sigma$ errors, or in one case slightly below, the "equilibrium value" from \cite{Mehta2023}. 
These equilibrium values are derived from stellar population synthesis models that suggest the ratio reaches an equilibrium value of $-1.93$ to $-1.78$  after 100-200~Myr of constant SFR for the metallicity range $\log(Z/${\Zsun}$)= [-2, 0]$. 
In open grey circles, we show the JADES galaxies at the same redshift range as the REBELS-IFU galaxies. 
Using this diagnostic ratio, \cite{Clarke2024} finds no trend in the burstiness of galaxies with redshift ($1.4<z<7$) and suggests that there is a mixture of bursty and non-bursty galaxies at all redshifts.

Whilst \cite{Sparre2017} and \cite{FloresVelazquez2021} argue that the H$\alpha$ to rest-UV flux ratio is a good indicator of star formation burstiness according to the FIRE simulations, other studies, such as \cite{Rezaee2023} in typical star-forming galaxies at $z\simeq2$ from the MOSDEF survey, cast doubt on the reliability of this ratio for probing bursty star formation.  
Using a spatially resolved analysis of the star formation surface density, stellar population ages, and spectral features in the rest-frame far UV, they find no evidence that elevated H$\alpha$-to-UV ratios are indicative of galaxies undergoing bursts of star formation.

The H$\alpha$-to-UV flux ratio is conventionally assumed to be a proxy for the 10-to-100~Myr SFR ratio.
In the right panel of Fig.~\ref{fig:burstiness}, we show that the SFR$_{10~\text{Myr}}$/SFR$_{100~\text{Myr}}$ ratios are all elevated above the value for a constant SFH and anticorrelate with mass (hinting at a possible mass dependence to the SFH burstiness, although the trend is not obvious in the JADES galaxies).
Similarly, we see that the JADES galaxies are also systematically shifted upwards when using this ratio.  
This is consistent with our model results shown in Fig.~\ref{fig:SFH_model}.
Thus, we show that the H$\alpha$-to-rest-UV ratio can be consistent with non-bursty values even when there is strong evidence for rapidly rising SFHs over the previous $100$~Myr (see Section~\ref{sec:discussion_NP_SFHs}).  
This can also be seen in the simulations of \cite{Asada2023} and \cite{Mehta2023}. 
Thus, we conclude that the H$\alpha$-to-UV ratio is not a completely reliable probe of burstiness and, at best, only provides a lower limit on how many galaxies are bursty \citep[e.g.][]{Clarke2024}.  
While some studies \citep[e.g.][]{Endsley2024, Kokorev2025} define bursty galaxies as those with $\log($SFR$_{10~\text{Myr}}$/SFR$_{100~\text{Myr}})>0$, \cite{Carvajal-Bohorquez2025} suggests only galaxies with values greater than $\gtrsim0.2$ should be considered bursty. 
Even with this more stringent constraint, the REBELS-IFU galaxies would be classified as bursty.  
In addition to the problems introduced by SFH variations, we note that observed H$\alpha$-to-rest-UV ratios are highly dependent on dust attenuation correction assumptions, since both the numerator and denominator require correction, and that the H$\alpha$-to-UV ratio predicted for a constant SFH is influenced by whether binary stellar evolution is included in the models and the choice of IMF \citep[see][]{Rezaee2023}.
Therefore, we suggest that SED fitting is likely to be a more reliable method of identifying bursty SFHs, including tests such as inspecting whether features in the galaxy spectra can be reproduced with constant SFHs.  
We also note that we have not considered shorter timescale (stochastic) fluctuations in SFHs.  This is beyond the scope of this work, but we note that stochastic fluctuations superimposed on a rising SFH \citep[e.g.][]{Kohandel2025} would be hard to distinguish from a steeply rising SFH.

\section{Conclusions}
\label{sec:summary} 
In this work, we present a comprehensive analysis of a range of SFR tracers from a sample of 12 massive Lyman-break galaxies at redshifts $z\simeq6.5-7.7$ known as the REBELS-IFU sample.  
We derive SFRs from the H$\alpha$, rest-UV, and FIR emission and compare these to those obtained from SED fits to the NIRSpec spectra.
The main conclusions of this work are:

\begin{itemize}
    \item We find an average stellar-to-nebular attenuation ratio of $f = E(B-V)_{\text{stellar}}/E(B-V)_{\text{gas}} = 0.50 \pm 0.08$, indicating that differential attenuation remains significant in massive, $z\simeq7$ galaxies. This ratio is consistent with the $f = 0.44$ factor for local star-forming galaxies \citep{Calzetti1997}, suggesting no clear redshift evolution.  
    However, the large scatter around the average relation implies that nebular attenuation cannot be reliably inferred from the stellar continuum for individual galaxies. 
    We find tentative evidence that the $f$ ratio is metallicity dependent, with the most metal-rich galaxy (REBELS-29) having a ratio consistent with $f=1$.  
    \item The average attenuation curve inferred from the {\irxb} relation lies between the Calzetti-like and SMC relations, consistent with the attenuation curves derived directly from the NIRSpec rest-UV/optical spectra in \cite{Fisher2025}.
    The REBELS-IFU galaxies exhibit high obscured SFR fractions, with $f_{\text{obs}} = $~SFR$_{\text{IR}}$/SFR$_{\text{UV+IR}} = 0.56 - 0.78$, and we utilise the multi-wavelength observations available for our sample to correct the dust-attenuated SFR tracers.
    \item The REBELS-IFU galaxies lie systematically above literature $z=7$ star-forming main-sequence relations when using the total SFRs derived from the rest-UV and FIR fluxes (SFR$_{\text{UV+IR}}$), the rest-UV SFRs from the reconstructed dust-free intrinsic SED (SFR$_{\text{UV, intrinsic SED}}$), the dust-corrected H$\alpha$ SFRs (SFR$_{\text{H}\alpha}$), or the SFRs averaged over a $10$~Myr timescale from the SED fits (SFR$_{10~\text{Myr}}$).  
    However, the SFRs averaged over a $100$~Myr timescale derived from the SED fits (SFR$_{100~\text{Myr}}$) are consistent with literature star-forming main-sequence relations.
    These discrepancies can be explained with rising SFHs, which causes SFR$_{\text{UV+IR}}$ (calculated using the fiducial luminosity-to-SFR conversion factors that are derived assuming a constant SFH) to match the SFR averaged over a timescale of $\sim20$~Myr, not the assumed $100$~Myrs. 
    Thus, we demonstrate that it can be misleading to adopt standard SFR conversion factors in high-redshift galaxies that are more likely to have rising or bursty SFHs, especially when comparing SFR$_{\text{UV+IR}}$ to SFR$_{\text{100 Myr}}$ or lower-redshift results.
    \item We find strong evidence for rising SFHs in the REBELS-IFU galaxies when fitting SED models to the integrated spectra.
    Compared to galaxies from the JADES survey at similar redshifts, the average REBELS-IFU SFH rises more steeply, which is surprising given their high stellar masses, but can be understood considering the rest-UV bright selection of our sample. 
    \item 
    Assuming the best-fit non-parametric SFHs for the REBELS-IFU galaxies, we provide new luminosity-to-SFR calibrations more relevant for obtaining the SFRs averaged over $100$~Myr of massive $z\simeq7$ galaxies.
    We find a median rest-UV luminosity to SFR$_{100~\text{Myr}}$ conversion factor of 
    \begin{eqnarray}
        \kappa_{\text{UV}}=(2.7\pm0.9) \times 10^{-29}~{\text{\Msun}}~\text{yr}^{-1}~\text{erg}^{-1}~\text{s Hz}
    \end{eqnarray}
    and a median H$\alpha$ luminosity to SFR$_{10~\text{Myr}}$ conversion factor of 
    \begin{eqnarray}
        \kappa_{\text{H}\alpha}=(4.7\pm1.4)~\times~10^{-42}~{\text{\Msun}}~\text{yr}^{-1}~\text{erg}^{-1}~\text{s}.
    \end{eqnarray}
    The spread in these values reflects at least a 30 per cent systematic uncertainty due to factors such as SFH shape and metallicity.
    The significant systematic decrease in $\kappa_{\text{UV}}$ by a factor of $\simeq 3$ ($>0.4$~dex) compared to the fiducial value derived assuming a constant SFH is of particular importance for the reliability of SFRs of $z>12$ objects, where observations typically only probe the rest-UV and SFHs are more likely to be rising.  
    Using model galaxy SEDs from BAGPIPES, we find $\kappa_{\text{UV}}$ and $\kappa_{\text{H}\alpha}$ both decrease with more steeply rising SFHs, with $\kappa_{\text{UV}}$ being more sensitive.  
    Therefore, to avoid overestimating the average SFR over a given timescale in high-redshift galaxies with rising SFHs, adjusted conversion factors should be used.
    In Fig.~\ref{fig:SFH_model}, we provide a graph to estimate the shift in conversion factors for a given SFH steepness. 
    \item Despite all the REBELS-IFU galaxies having SFR$_{10~\text{Myr}}$/SFR$_{100~\text{Myr}}>1$, indicative of bursty SFHs, the observed H$\alpha$-to-UV luminosity ratios for all but one of the galaxies remain consistent (within $1\sigma$) with the non-bursty value. 
    This result is supported by our BAGPIPES model spectra that show the SFR$_{\text{H}\alpha}$/SFR$_{\text{UV}}$ ratio varies to a lesser degree than than the SFR$_{10~\text{Myr}}$/SFR$_{100~\text{Myr}}$ ratio with increasing SFH steepness due to the sensitivity of the rest-UV flux to fluctuations in SFR on timescales shorter than $100$~Myr.  
    This suggests the H$\alpha$-to-UV luminosity ratio is not a reliable indicator of bursty SFHs, even with robust dust attenuation corrections.  
\end{itemize}

We have shown that using multi-wavelength data provides a more comprehensive picture of the global properties of these galaxies.  
In future work, we will extend this analysis to spatially resolved scales.  
This is now possible given the wealth of multi-wavelength data we now have for some of these sources - including NIRCam imaging (PID 6480; P.I. Schouws, PID 6036; P.I. Hodge), the NIRSpec IFU spectroscopy, and high-resolution (up to $0.15''$) ALMA Band 6 data for ten of the REBELS-IFU galaxies \citep[see][for REBELS-25 and Phillips et al. in prep. and the recently approved ALMA PID: 2025.1.01318.S, P.I. Stefanon]{Rowland2024}. 
This will further reveal how these galaxies build up their dust and stellar mass in the first Gyr of the Universe.

\section*{Acknowledgements}
We thank the authors of Tsujita et al. 2025 and Faisst et al. 2025 for sharing their ALPINE results with us pre-publication.  

RB acknowledges support from an STFC Ernest Rutherford Fellowship [grant number ST/T003596/1].
RKC is grateful for support from the Leverhulme Trust via the Leverhulme Early Career Fellowship.

MA is supported by FONDECYT grant number 1252054, and gratefully acknowledges support from ANID Basal Project FB210003 and ANID MILENIO NCN2024\_112.
PD warmly acknowledges support from an NSERC discovery grant (RGPIN-2025-06182).
JAH acknowledges support from the ERC Consolidator Grant 101088676 (“VOYAJ”).
MS acknowledges support from the European Research Commission Consolidator Grant 101088789 (SFEER), from the CIDEGENT/2021/059 grant by Generalitat Valenciana, and from project PID2023-149420NB-I00 funded by MICIU/AEI/10.13039/501100011033 and by ERDF/EU.

This work is based on observations made with the NASA/ESA/CSA \textit{James Webb Space Telescope}. The data were obtained from the Mikulski Archive for Space Telescopes at the Space Telescope Science Institute, which is operated by the Association of Universities for Research in Astronomy, Inc., under NASA contract NAS 5-03127 for \textit{JWST}. These observations are associated with programs \#1626 and \#2659.

\section*{Data Availability}
The data used in this manuscript will be made available to others upon reasonable request to the authors.



\bibliographystyle{mnras}
\bibliography{export25Jun25} 




\appendix

\section{The effect of attenuation curve assumptions on nebular-to-stellar attenuation ratios}
\label{sec:EBV_different_neb_curve}
\begin{figure*} 
\includegraphics[width=\columnwidth]{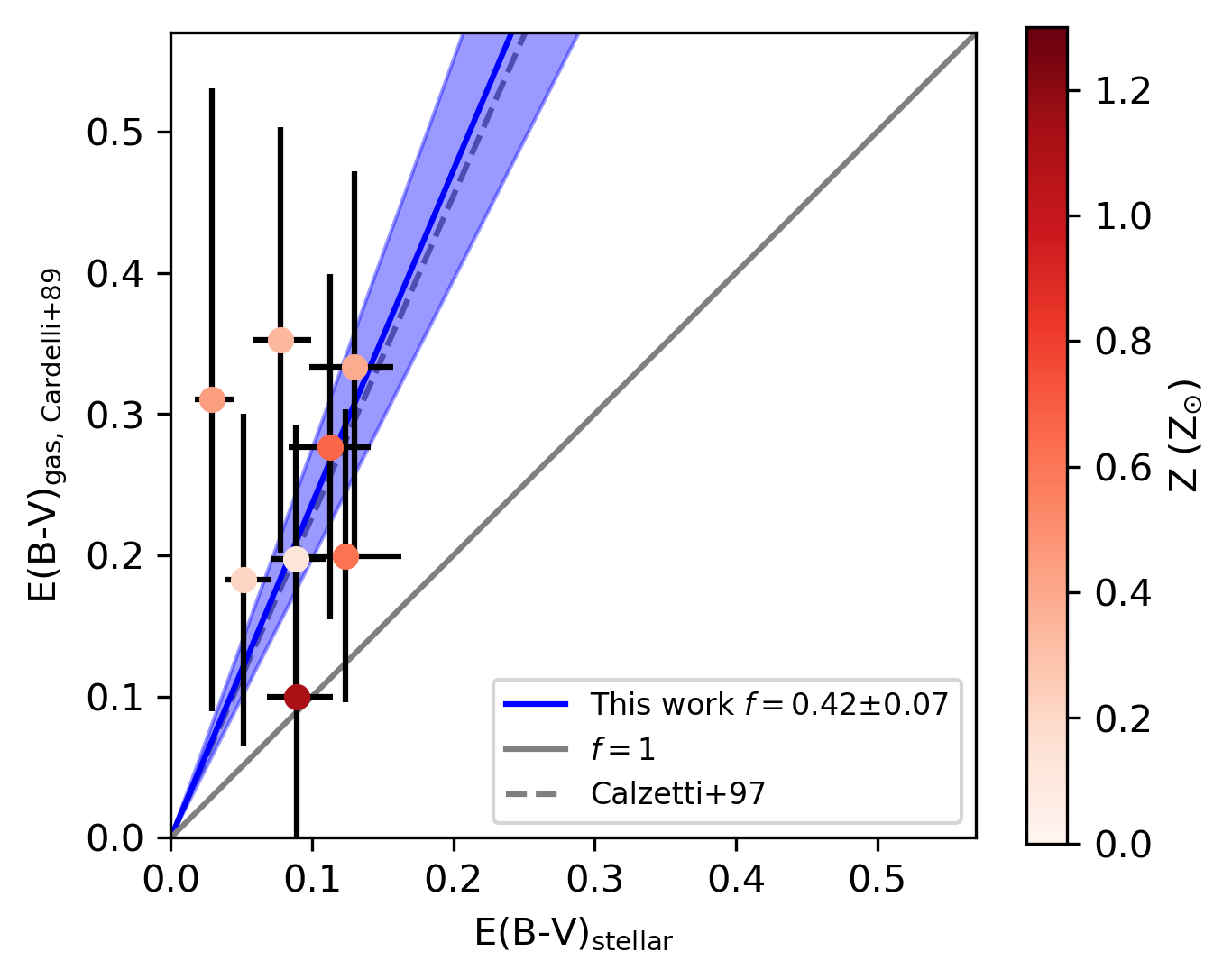}
\includegraphics[width=\columnwidth]{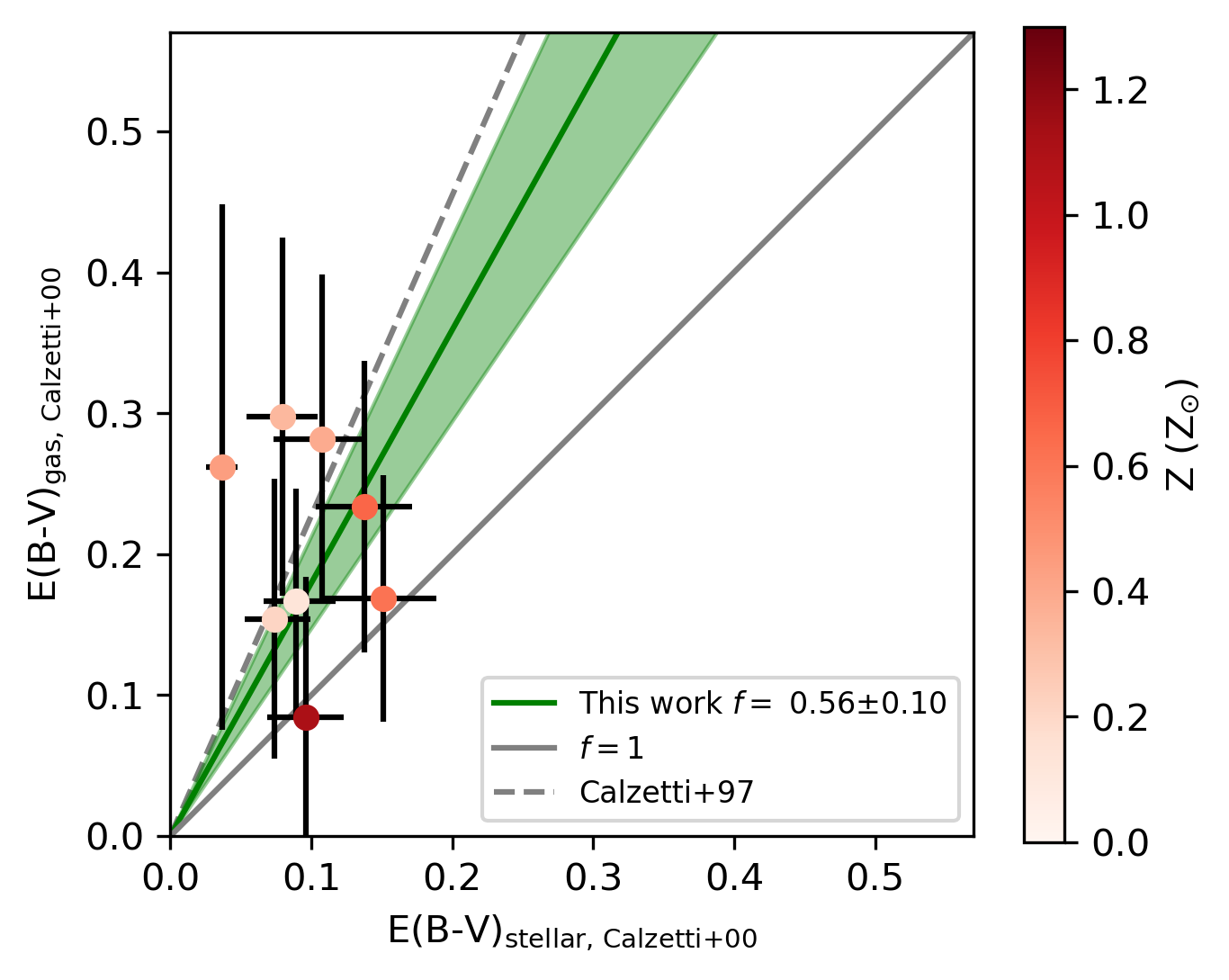}
\caption{
The colour excesses for the ionised gas, $E(B-V)_{\text{gas}}$, derived from the Balmer decrement, compared to the colour excesses for the stellar continuum, $E(B-V)_{\text{stellar}}$, assuming different attenuation (or extinction) curves.
In the left panel, we obtain a slightly steeper gradient when assuming the \citet{Cardelli1989} extinction curve for $E(B-V)_{\text{gas}}$ and the attenuation curves from \citet{Fisher2025} for $E(B-V)_{\text{stellar}}$.
In the right panel, we obtain a slightly shallower gradient when assuming the \citet{Calzetti2000} attenuation curve for both $E(B-V)_{\text{gas}}$ and $E(B-V)_{\text{stellar}}$.
However, both gradients are consistent,` within the errors, with our fiducial fit shown in Fig.~\ref{fig:EBV_gas_vs_stellar}.}
\label{fig:EBV_gas_vs_stellar_appendix} 
\end{figure*}

In Fig.~\ref{fig:EBV_gas_vs_stellar_appendix} we show the colour excesses for the ionised gas, $E(B-V)_{\text{gas}}$, derived from the Balmer decrement, compared to the colour excesses for the stellar continuum, $E(B-V)_{\text{stellar}}$, with different attenuation curve assumptions.
In the left panel we assume the \cite{Cardelli1989} extinction curve for $E(B-V)_{\text{gas}}$ and the attenuation curves from \cite{Fisher2025} for $E(B-V)_{\text{stellar}}$.
The average ratio is slightly steeper but consistent within the errors with our fiducial fit shown in Fig.~\ref{fig:EBV_gas_vs_stellar}. 
In the right panel we show that assuming the \cite{Calzetti2000} attenuation curve for both $E(B-V)_{\text{gas}}$ and $E(B-V)_{\text{stellar}}$ results in a slightly shallower slope, but it is still consistent within the errors with our fiducial fit. 
Thus, we conclude that our attenuation curve assumptions do not significantly affect our main results.

\section*{Affiliations}
\noindent
{\it
$^{1}$Jodrell Bank Centre for Astrophysics, University of Manchester, Oxford Road, Manchester M13 9PL, UK\\
$^{2}$Leiden Observatory, Leiden University, P.O. Box 9513, 2300 RA Leiden, The Netherlands \\
$^{3}$Departament d’Astronomia i Astrofìsica, Universitat de València, C. Dr. Moliner 50, E-46100 Burjassot, València, Spain \\
$^{4}$Unidad Asociada CSIC ”Grupo de Astrofísica Extragaláctica y Cosmología” (Instituto de Física de Cantabria - Universitat de València), Spain \\
$^{5}$Institute of Astronomy and Astrophysics, Academia Sinica, 11F of Astronomy-Mathematics Building, No.1, Sec. 4, Roosevelt Rd, Taipei 106216, Taiwan, R.O.C. \\
$^{6}$Hiroshima Astrophysical Science Center, Hiroshima University, 1-3-1 Kagamiyama, Higashi-Hiroshima, Hiroshima 739-8526, Japan \\
$^{7}$National Astronomical Observatory of Japan, 2-21-1, Osawa, Mitaka, Tokyo, Japan \\
$^{8}$ Instituto de Estudios Astrof\'{\i}cos, Facultad de Ingenier\'{\i}a y Ciencias, Universidad Diego Portales, Av. Ej\'ercito 441, Santiago, Chile\\
$^{9}$ Millenium Nucleus for Galaxies (MINGAL)\\
$^{10}$ International Centre for Radio Astronomy Research, University of Western Australia, 35 Stirling Hwy., Crawley, WA 6009, Australia\\
$^{11}$ Research School of Astronomy and Astrophysics, Australian National University, Canberra, ACT 2611, Australia\\
$^{12}$ ARC Centre of Excellence for All Sky Astrophysics in 3 Dimensions (ASTRO 3D), Australia\\
$^{13}$ Canadian Institute for Theoretical Astrophysics, 60 St George St, University of Toronto, Toronto, ON M5S 3H8, Canada \\
$^{14}$ David A. Dunlap Department of Astronomy and Astrophysics, University of Toronto, 50 St George St, Toronto ON M5S 3H4, Canada \\
$^{15}$ Department of Physics, 60 St George St, University of Toronto, Toronto, ON M5S 3H8, Canada\\
$^{16}$Scuola Normale Superiore, Piazza dei Cavalieri 7, 56126 Pisa, Italy \\
$^{17}$Astrophysics Research Institute, Liverpool John Moores University, 146 Brownlow Hill, Liverpool L3 5RF, UK\\
$^{18}$Center for Computational Astrophysics, Flatiron Institute, 162 5th Avenue,
New York, NY 10010, USA \\
$^{19}$ Department of Astronomy, University of California, 501 Campbell Hall \#3411, Berkeley, CA 94720, USA\\
}

\bsp	
\label{lastpage}
\end{document}